\documentclass[12pt]{article}

\usepackage{amsmath,amsfonts,bm}

\def\1{\bm{1}}

\def\vzero{{\bm{0}}}

\def\ve{{\bm{e}}}

\def\mC{{\bm{C}}}

\def\mN{{\bm{N}}}

\def\mY{{\bm{Y}}}

\DeclareMathAlphabet{\mathsfit}{\encodingdefault}{\sfdefault}{m}{sl}
\SetMathAlphabet{\mathsfit}{bold}{\encodingdefault}{\sfdefault}{bx}{n}

\def\gB{{\mathcal{B}}}

\def\gF{{\mathcal{F}}}

\def\gH{{\mathcal{H}}}

\def\gN{{\mathcal{N}}}

\def\gP{{\mathcal{P}}}

\def\sN{{\mathbb{N}}}

\def\sT{{\mathbb{T}}}
\def\sU{{\mathbb{U}}}

\newcommand{\E}{\mathbb{E}}

\DeclareMathOperator*{\argmin}{arg\,min}

\usepackage{amssymb,amsmath,amsthm,amsfonts,eurosym,geometry,ulem,graphicx,caption,color,setspace,sectsty,comment,footmisc,caption,natbib,pdflscape,subfigure,array,hyperref}
\usepackage{float}
\usepackage{url}

\usepackage{todonotes}

\usepackage{accents}
\newcommand{\ubar}[1]{\underaccent{\bar}{#1}}

\newcommand{\be}{\begin{equation}}
\newcommand{\bi}{\begin{itemize}}
\newcommand{\ee}{\end{equation}}
\newcommand{\ei}{\end{itemize}}
\newcommand{\bea}{\begin{eqnarray*}}
\newcommand{\eea}{\end{eqnarray*}}
\newcommand{\bean}{\begin{eqnarray}}
\newcommand{\eean}{\end{eqnarray}}

\newtheorem{theorem}{Theorem}

\newtheorem{proposition}{Proposition}
\newtheorem{proofTheorem}[proposition]{Sketchy Proof of Theorem}

\newcolumntype{L}[1]{>{\raggedright\let\newline\\arraybackslash\hspace{0pt}}m{#1}}
\newcolumntype{C}[1]{>{\centering\let\newline\\arraybackslash\hspace{0pt}}m{#1}}
\newcolumntype{R}[1]{>{\raggedleft\let\newline\\arraybackslash\hspace{0pt}}m{#1}}

\begin{document}

\begin{titlepage}
\title{Inference for Model Misspecification in Interest Rate Term Structure using Functional Principal Component Analysis}
\author{
Kaiwen Hou\thanks{Columbia Business School, \href{mailto:kaiwen.hou@columbia.edu}{kaiwen.hou@columbia.edu}}
}
\date{\today}
\maketitle
\begin{abstract}
\noindent 
Level, slope, and curvature are three commonly-believed principal components in interest rate term structure and are thus widely used in modeling. This paper characterizes the heterogeneity of how misspecified such models are through time. Presenting the orthonormal basis in the Nelson-Siegel model interpretable as the three factors, we design two nonparametric tests for whether the basis is equivalent to the data-driven functional principal component basis underlying the yield curve dynamics, considering the ordering of eigenfunctions or not, respectively. Eventually, we discover high dispersion between the two bases when rare events occur, suggesting occasional misspecification even if the model is overall expressive.
\\
\vspace{0in}
\vspace{0in}

\bigskip
\end{abstract}
\setcounter{page}{0}
\thispagestyle{empty}
\end{titlepage}
\pagebreak \newpage

\doublespacing

\section{Introduction} \label{sec:introduction}
The dynamic evolution of the interest rate term structure is indispensable for many tasks, including pricing interest rate derivatives, discounting future payoff for asset prices, explaining return predictability, conducting firm evaluations, managing financial risk, making monetary and fiscal policy~\citep{elenev2021can,elenev2022can,krishnamurthy2011effects,woodford2012methods}, and most importantly, forecasting rare events like financial crises and predicting future growth~\cite{hu2013noise,koijen2017cross}. 

\begin{figure}[H]
    \centering
    \includegraphics[width=\textwidth]{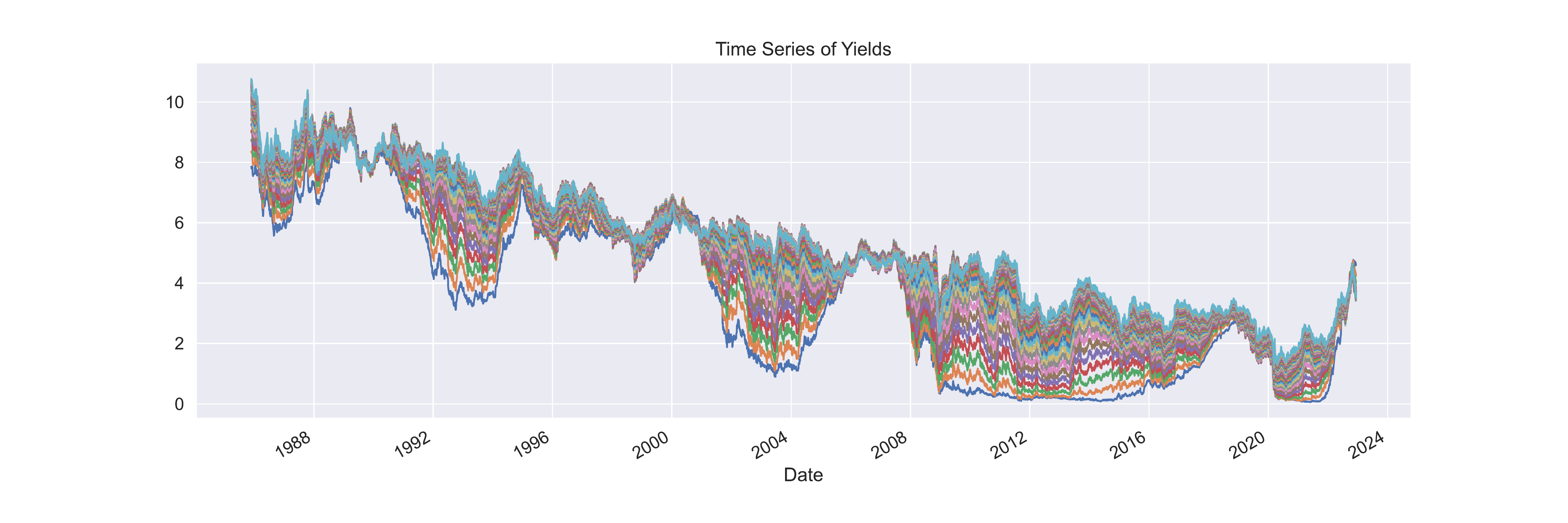}
    \caption{Time series of zero-coupon yields on different bond maturities shows heteroscedastic covariation of yield curves.}
    \label{fig:timeSeriesYields}
\end{figure}

\begin{figure}[H]
    \centering
    \includegraphics[width=\textwidth]{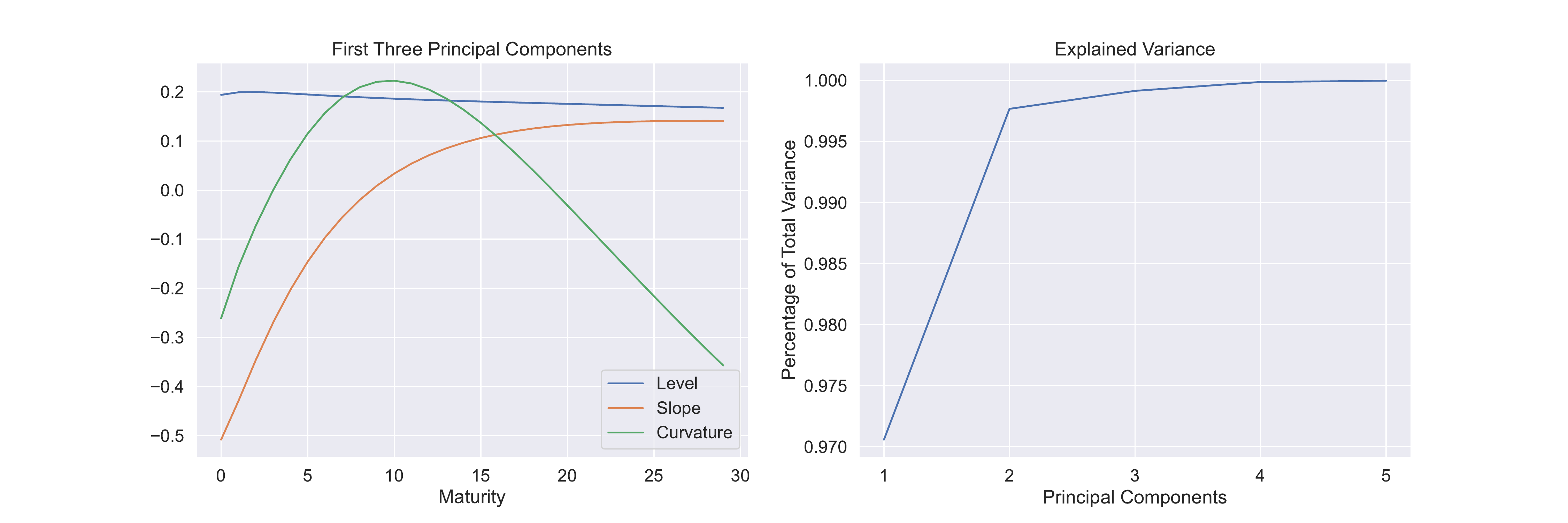}
    \caption{First three principal components of the entire dataset and explained variance ratios.}
    \label{fig:first3PC_allData}
\end{figure}

There \textit{seems} to be a linear structure in the interest rate term structure \textit{most of the times}. Eigendecomposition of the covariance matrix of yield changes attributes over 97\% of the
variance to three principal components, known as the level, the slope, and the curvature, seemingly universal across different units of maturity~\citep{chapman2001recent,litterman1991common,piazzesi2010affine}. Using the daily zero-coupon yield curve of bonds maturing in 1 year to 30 years~\citep{gurkaynak2007us} from November 25, 1985\footnote{The data starts in 1985 after dropping the missing and incomplete data, so that all cross-sectional data points have observations from hypothetical securities with 1-year to 30-year maturity. The missing data before was mainly due to lack of actual securities for model calibration.}, to December 9, 2022, we obtain the first three principal components of the entire dataset in Figure~\ref{fig:first3PC_allData}.

\begin{figure}[H]
    \centering
    \includegraphics[width=\textwidth]{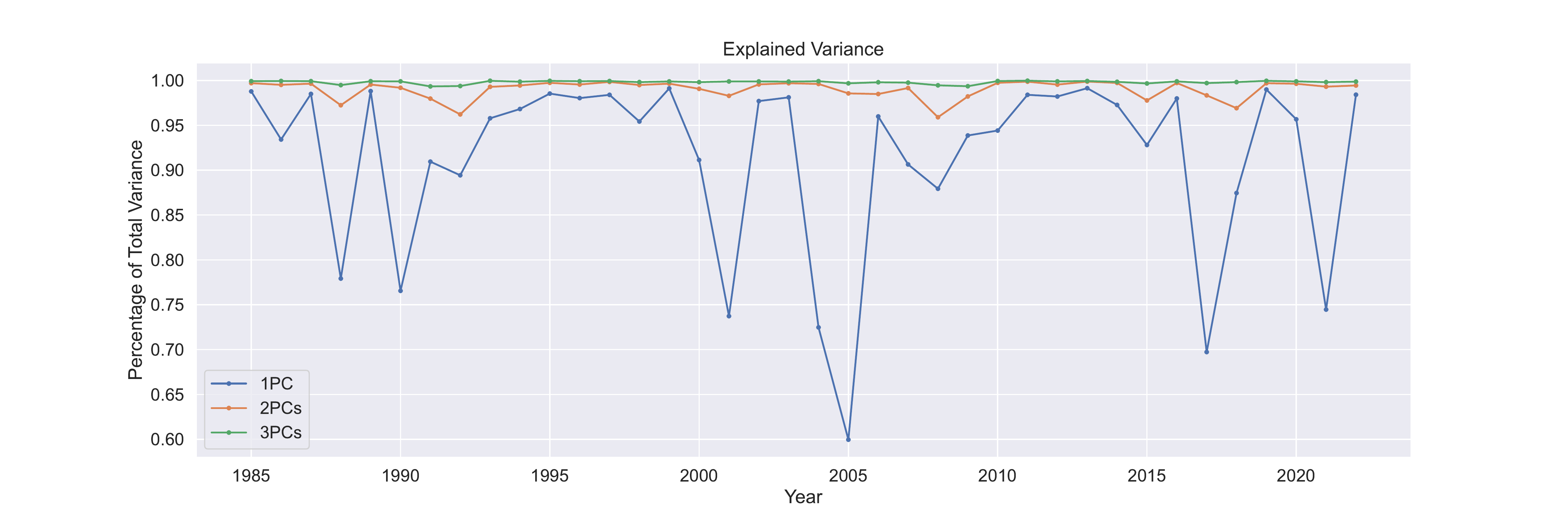}
    \caption{Explained variance by first 1/2/3 principal components in each year.}
    \label{fig:explainedVar}
\end{figure}

On the other hand, a dispersion in actual yields around the smooth term structure has been used to measure liquidity, whose friction spikes in crises while remaining small otherwise~\citep{hu2013noise}. 
Following~\citet{fama1973risk} regression and construction of a factor mimicking portfolio~\citep{ang2006cross}, a significant liquidity premium is revealed so that the noise measure passes the cross-sectional test. However, the decent returns gained from trading illiquidity do not necessarily suggest that the interest-rate model is well-specified, nor does it justify using a smooth three-factor term structure model as a reference model (as in~\citet{hu2013noise}) from which dispersion is measured. 

\begin{figure}[H]
    \centering
    \includegraphics[width=\textwidth]{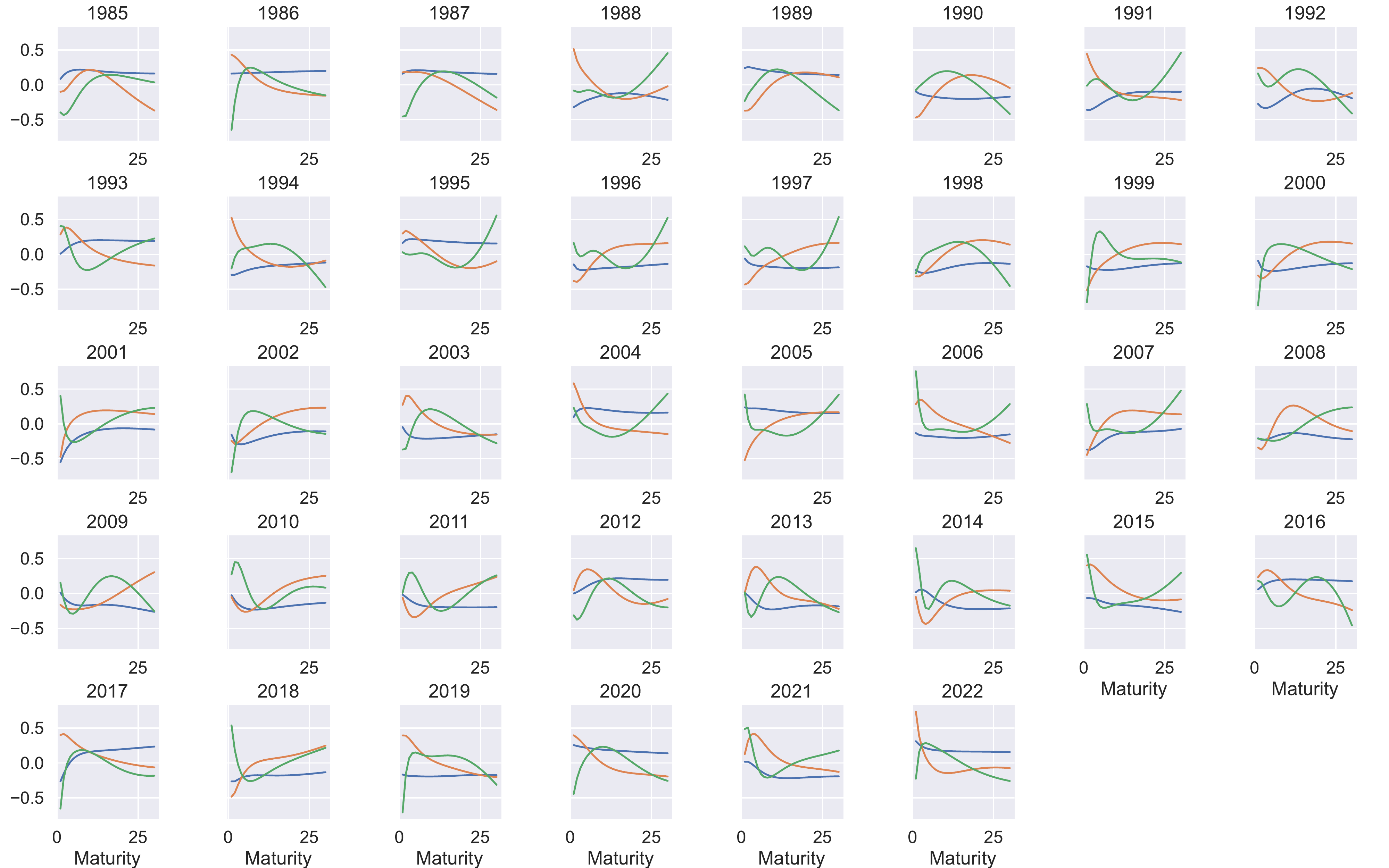}
    \caption{Loadings of first three principal components in each year. Blue curves represent the first principal component or the ``level" factor; orange curves represent the second principal component or the ``slope" factor; green ones represent the third principal component or the ``curvature".}
    \label{fig:first3PC_year}
\end{figure}

To put it in another way, even if the first three principal components can explain a high variance in the yield curves (which also means that reconstruction error is low), as shown in Figure~\ref{fig:explainedVar}, and even if the full-sample stylistic features (in Figure~\ref{fig:first3PC_allData}) of first three principal components have an immaculate interpretation of their shapes, the three principal components do not always correspond to the level, slope, and curvature factors from a more granular perspective - in some years they may not have rich economic interpretations at all~\citep{lord2007level}.
Indeed, from Figure~\ref{fig:first3PC_year}, the level-slope-curvature seems to be occasionally violated, as sometimes the first principal component (which ``ought to be" the level) is not flat (e.g., in 2007 and 2017), and the second principal component (which ``ought to be" the slope) is not monotonic (e.g., in 1990, 2008, and 2021).

With such economic interpretations being invalidated occasionally, some standard models of interest rate term structure based on these interpretations, such as~\citet{nelson1987parsimonious}, might problematically impose a specific smooth structure of yield curves, and the dispersion from a smooth yield curve that ~\citet{hu2013noise} benchmarks against could be a biased benchmark.

As pointed out in previous literature, model misspecification (also known as model uncertainty or ambiguity) should be acknowledged in macroeconomic theory~\citep{hansen2001acknowledging,hansen2001robust}; market participants probably have a preference towards robustness~\citep{anderson2000robustness,hansen1999robust,maenhout2004robust}; the monetary policies could have differed a lot when the actual model describing the environment appears ambiguous to the central bank~\citep{giannoni2002does,hansen2010wanting,kasa2002model,onatski2002robust,woodford2010robustly}. It is also likely that interest rate models also suffer from misspecification~\citep{brenner1996another}, mainly when market crashes or jumps are observed in continuous time~\citep{johannes2004statistical}. Despite the overall satisfactory exponentially affine term structure models, the moments of latent variables present in the affine structure could also be ambiguous~\citep{liu2018return,liu2022volatility,zhang2022optimal}, leading to a misspecified affine model.

\begin{figure}[H]
    \centering
    \includegraphics[width=\textwidth]{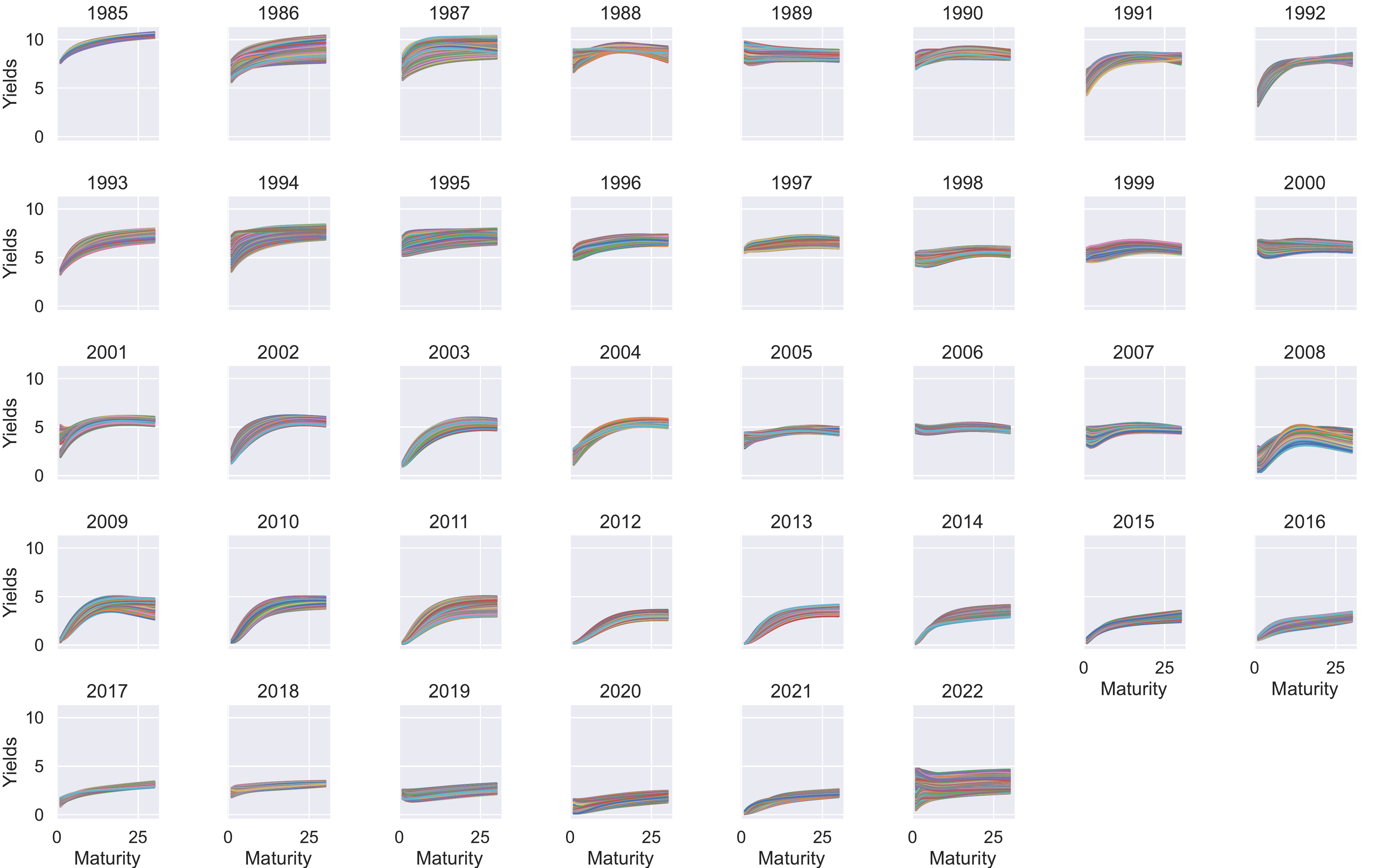}
        \caption{Daily yield curves from 1985 to 2022.}
    \label{fig:yieldCurveYear}
\end{figure}

In light of~\citet{hu2013noise}, this paper examines the model misspecification of interest rate term structures both on an omniscient scope and a granular perspective. Figure~\ref{fig:yieldCurveYear} presents the functional data structure we have, i.e., the time series of term structure as a function, which motivates our methodology different from the inadequate conventional one.
We utilize recent advances in functional data analysis, particularly a nonparametric hypothesis test of functional principal component analysis (FPCA), and compare its results implied by classical principal component analysis (PCA), which highly relies on the homogeneity of term structure throughout time. Instead of directly constructing a stochastic partial differential equation to model the interest rate dynamics as a stochastic process in the space of functions~\citep{cont2005modeling}, we employ the nonparametric and distribution-free method in this research to lean on minimal assumptions.

Our codes are released on \url{https://github.com/Kaiwen-Hou-KHou/termStructureFPCA}.

\section{Preliminaries}
In this section, we introduce the essential statistical and econometric tools to deal with functional data such as the time series of yield curves, including a generalization of classical principal component analysis and eigendecomposition into infinite-dimension space $L^2([0,1],d\xi)$ (the space of square-integrable functions on $[0,1]$, where the inner product weighted by the constant function $1$ is defined as $\langle f,g \rangle := \int\limits_0^1 fg d\xi$), the assumption of smoothness and smoothing techniques, followed by some asymptotics that helps examine FPCA method through statistical inference.

\subsection{Functional Principal Component Analysis}
A generalized PCA method, FPCA is developed to investigate the dominant modes of functional data variation, such as modeling sparse longitudinal data or conducting functional classification and regression~\citep{hall2006properties,li2010uniform,yao2005functional,yao2005functional2}.

Let $y_i$ be i.i.d. samples of a random function $y: (\Omega, \gF, P)\to (C^0[0,1], \gB_{C[0,1]})$ with $\E||y||^2<\infty$. Denote by 
$$
m=\E y\in L^2([0,1],d\xi)
$$ 
the mean function, and by 
$$
G(x,x')=Cov(y(x),y(x'))\in L^2([0,1]^2,d\xi d\xi')
$$ 
the covariance function\footnote{Since covariance is symmetric (and continuous here), then $G$ is a positive semi-definite kernel, so Mercer's theorem applies.}. By Mercer's theorem, there exists a sequence of eigenvalues $\lambda_k \downarrow 0$ such that $\sum\limits_{k=1}^\infty \lambda_k <\infty$ and
$$
G(x,x') = \sum\limits_{k=1}^{\infty}\lambda_k\psi_k(x)\psi_k(x'),
$$
where the orthonormal basis functions $(\psi_k)_{k=1}^\infty$ are the eigenfunctions satisfying
$$
\int\limits_0^1 G(x,x')\psi_k(x')dx' = \lambda_k \psi_k(x).
$$
By the Kosambi–Karhunen–Lo\`eve theorem~\citep{kosambi2016statistics}, one can write 
$$
\xi_i = y_i - m = \sum\limits_{k=1}^\infty \zeta_{ik}\psi_k,
$$
where the principal component\footnote{$\zeta_{ik}$ is also known as the rescaled functional principal component scores.} associated with the $k$-th eigenfunction $\psi_k$ is $\zeta_{ik} = \langle \xi_i, \psi_k \rangle$ with the following moment properties 
\bea
\E\zeta_{ik} &=& 0, \\
\E\zeta_{ik}^2 &=& \lambda_k, \\
\E\zeta_{ik}\zeta_{ik'} &=& 0,
\eea
for all $1\leq i\leq n,$ and $ k\neq k'\in\sN^+$.

It is also justified in \citet{cont2005modeling} and \citet{kusuoka2000term}\footnote{This is a work suggested by Professor Darrell Duffie.} that the term structure could be modeled to have a two-dimensional dynamics (time and maturity), even if the dynamics on maturity is slightly anti-intuitive. Therefore, the interest rate term structure fits in this stochastic framework.

\subsection{Smoothness}
Note that a crucial assumption of FPCA is that data are modeled as sample paths of smooth stochastic processes, although they are often observed discretely and noisily. 
If sample paths are not smooth, then $G\notin L^2([0,1]^2,d\xi d\xi')$. Fortunately, this assumption can be relaxed by restricting $G$ to a subset to model some diffusion processes. For example, standard Brownian motion has non-differentiable covariance $G(x,x')=\min(x,x')$ on the diagonal, but it becomes $G(x,x')=x$, which is infinitely differentiable when restricted on the upper-triangular block
\be
\sU\sT:= \{(x,x')\mid 0\leq x < x' \leq 1 \}.
\label{eq:triangle} \ee
 With similar restriction, the covariance functions of common diffution processes such as Brownian bridge, geometric Brownian motion, and Ornstein-Uhlenbeck process~\citep{jouzdani2021functional} would also lie in $L^2([0,1]^2,d\xi d\xi')$.

For discrete observations, methods like local linear or spline smoothing can be used to obtain the functional form of \textit{each} $y_i$~\citep{yao2005functional}. 
In particular, if observations are evenly-spaced at $\left(\frac{j}{N}\right)_{j=1}^N$, one can adopt the B-spline estimator by projecting the discrete-valued function $y_i$ to the space $\gH^{(p-2)}[0,1]$ spanned by B-splines of order $p$:
\be
\hat{y}_i = \argmin_{g\in\gH^{(p-2)}[0,1]} \sum\limits_{j=1}^N 
\left\{ 
y_i\left(\frac{j}{N}\right) - g\left(\frac{j}{N}\right)
\right\}^2.
\label{eq:Bspline}\ee
Correspondingly, the smoothed and centered process is estimated by
$$
\hat{\xi}_i = \hat{y}_i - \hat{m} = \hat{y}_i - \frac{1}{n}\sum\limits_{j=1}^n \hat{y_j}.
$$
Getting rid of the diagonal components in $G$ by restrictions in Eq.~\eqref{eq:triangle} also ignores the measurement errors after smoothing~\citep{staniswalis1998nonparametric}.

In dealing with interest rate term structures, their smoothness has been a reasonable assumption as widely used in interpolating yield curves, which rely on piecewise smooth functions that are also smoothly joined at selected knots~\citet{christensen2011affine,diebold2006forecasting,diebold2008global,fisher1995fitting,mcculloch1975tax,nelson1987parsimonious,svensson1994estimating}.

\subsection{Hypothesis Testing}
Given an orthonormal set $(\psi_{0k})_{k=1}^{\infty}$, \citet{song2022hypotheses} propose a statistical test for the hypothesis $H_0$ that the eigenfunctions $(\psi_k)_{k=1}^\infty$ form the same set up to permutations and change of signs. Fixing a hyper-parameter $\kappa_n$, the test statistic is
\be
\hat{S}_n = n\sum\limits_{1\leq k< k'\leq\kappa_n} \hat{Z}_{kk'}^2,
\label{eq:testStat} \ee
where 
$$
\hat{Z}_{kk'} = \left\langle \hat{G}(x,x'), \psi_{0k}(x)\psi_{0k'}(x') \right\rangle_{L^2([0,1]^2,d\xi d\xi')} 
:= \int\limits_0^1\int\limits_0^1 \hat{G}(x,x')\psi_{0k}(x)\psi_{0k'}(x') dxdx',
$$
with the two-step estimate of the covariance
\be
\hat{G}(x,x') = \frac{1}{n}\sum\limits_{i=1}^{n}\hat{\xi}_i(x)\hat{\xi}_i(x').
\label{eq:covEstimate}\ee
Note that $\hat{G}$ is essentially a low-rank approximation of the bi-infinite covariance matrix, as $y_i$ has been projected onto $\gH^{(p-2)}[0,1]$. This na\"ive sub-block estimator is genuinely a choice out of simplicity; one could also adopt stochastic estimation such as the Russian roulette estimator to attain unbiasedness~\citep{hou2022spectral}.

It has also been shown in \citet{song2022hypotheses} that the asymptotic distribution of the test statistic under $H_0$ is an infinite Gaussian quadratic form, and a finite sample estimator of its quantile $\hat{Q}_{1-\alpha}$, which is the $100(1-\alpha)$-th percentile of
\be
\bar{S}_n = \sum\limits_{1\leq k< k'\leq\kappa_n} \hat{\lambda}_k\hat{\lambda}_{k'}\chi_{kk'}^2(1),
\label{eq:quadraticForm}\ee
is consistent. Here
$$
\hat{\lambda}_k = \frac{1}{n}\sum\limits_{i=1}^n \hat{\zeta}_{ik}^2,
$$
where
$$
\hat{\zeta}_{ik} = \left\langle \hat{\xi}_i,\psi_{0k} \right\rangle.
$$
The distribution of the Gaussian quadratic form $\bar{S}_n$ in Eq.~\eqref{eq:quadraticForm} is essentially a weighted sum of $\chi^2(1)$ random functions, where the weights are the eigenvalue pairs of $G$. In the special case where these eigenvalues are 1, the distribution becomes $\chi^2\left( \frac{\kappa_n(\kappa_n-1)}{2}\right)$, but this does not hold in general, as the scale parameter of a gamma distribution changes when multiplied by a constant and $\hat{\lambda}_k\hat{\lambda}_{k'}\chi_{kk'}^2(1)$ might be no longer of Chi-squared distribution. Unfortunately, $\bar{S}_n$ does not have a neat closed-form representation of its density or distribution function, so its quantiles should probably be computed numerically~\citep{bodenham2016comparison,davies1980algorithm}.

\section{Nelson-Siegel Model} \label{sec:NS}
The Nelson-Siegel model~\citep{nelson1987parsimonious} fits the yield curve with a simple functional form
\be
\hat{y}(\tau\mid \Theta) = \beta_0 + \beta_1 \frac{1-e^{-\theta\tau}}{\theta\tau}
+ \beta_2 \left(
\frac{1-e^{-\theta\tau}}{\theta\tau} - e^{-\theta\tau}
\right),
\label{eq:NSoriginal}\ee
where the parameters are $\Theta=(\beta_0, \beta_1, \beta_2, \theta)$ and $\tau$ is the maturity in months\footnote{The model is invariant with respect to the unit of maturity, as $\theta$ absorbs the scale difference between units. For example, transforming $\tilde{\theta}=12\theta$ and $\tilde{\tau}=\frac{\tau}{12}$ in years, Eq.~\eqref{eq:NSoriginal} is of the same form.}. 

It is acknowledged (e.g., by~\citet{diebold2006forecasting,diebold2008global}) that $\beta_0,\beta_1$, and $\beta_2$ represent loadings on the level factor, the slope factor, and the curvature factor, respectively, so Eq.~\eqref{eq:NSoriginal} can be generalized to the dynamic Nelson-Siegel Model in the following Eq.~\eqref{eq:NSdynamic}
\be
\hat{y}_t(\tau\mid \Theta) = L_t + S_t \frac{1-e^{-\theta\tau}}{\theta\tau}
+ C_t \left(
\frac{1-e^{-\theta\tau}}{\theta\tau} - e^{-\theta\tau}
\right),
\label{eq:NSdynamic}\ee
through adding the auto-regressive state variables $L_t$ for level, $S_t$ for slope, and $C_t$ for curvature.

We now proceed to test whether the orthonormal basis beneath the three factors $1, \frac{1-e^{-\theta\tau}}{\theta\tau}$ and $\left(
\frac{1-e^{-\theta\tau}}{\theta\tau} - e^{-\theta\tau}
\right)$ is equivalent to the actual principal component basis to see if the level, slope, and curvature factors always explain the major variation in the interest rate term structures, and offer insights on the dispersion between the two bases.

\subsection{Orthonormalization}
Typically the tradable bond maturity in months is bounded both from above and from below, i.e., $0<\ubar{\tau}\leq\tau\leq\bar{\tau}$. 
We can reproduce Eq.~\eqref{eq:NSoriginal} in the following theorem so that $\tau$ is rescaled in~$[0,1]$.
\begin{theorem}\label{thm:NS_ONB}
There exist three orthonormal basis functions $\psi_{0k}\in L^2([0,1],d\xi) \ (k=1,2,3)$ such that the Nelson-Siegel term structure $\hat{y}\in Span(\psi_{0k})_{k=1}^3 \subset L^2([0,1],d\xi)$.
\end{theorem}

\begin{figure}[H]
    \centering
    \includegraphics[width=\textwidth]{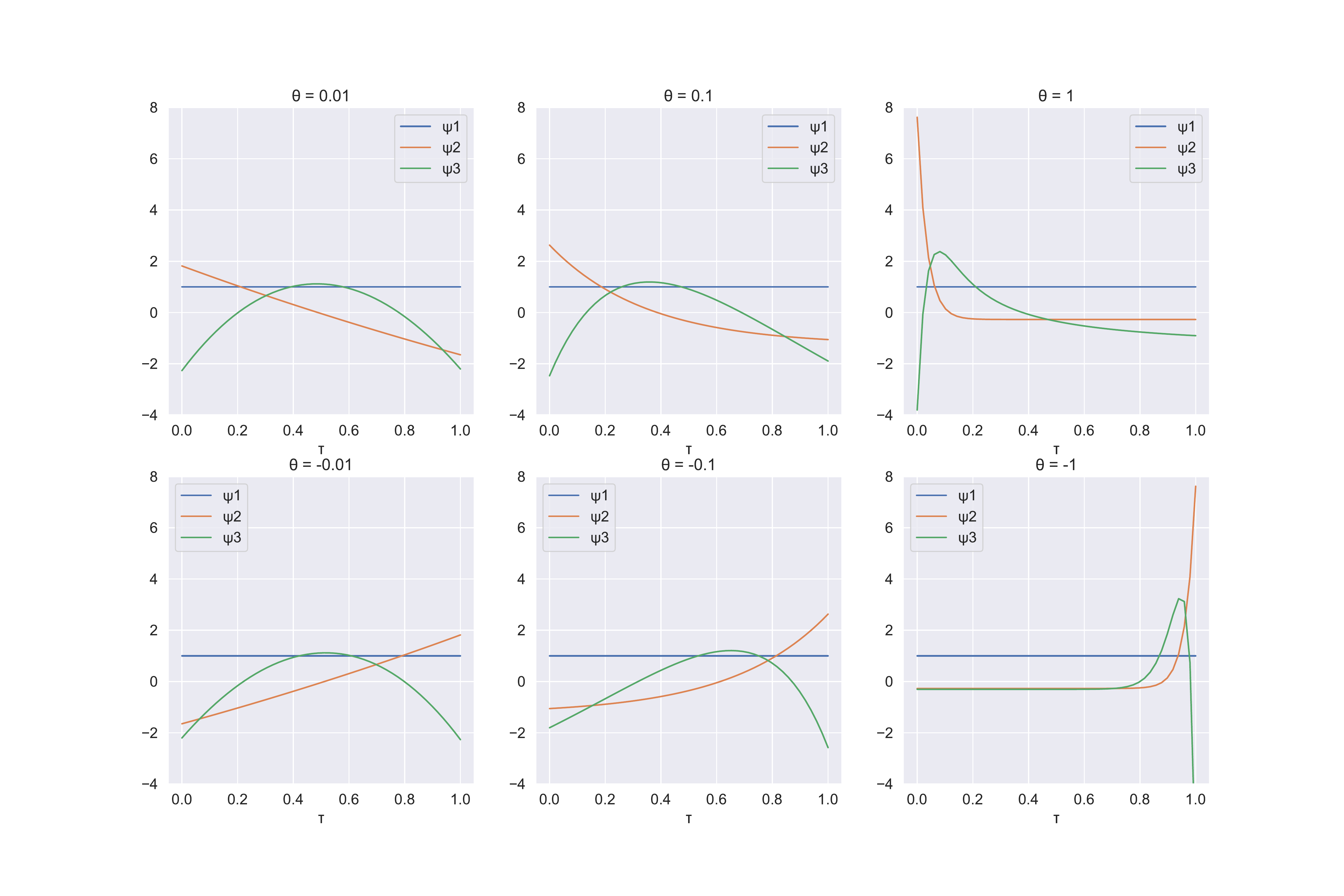}
    \caption{Orthonormal basis $(\psi_{0k})_{k=1}^3$.}
    \label{fig:NS_ONB}
\end{figure}

For the closed-form expressions of the orthonormal basis $(\psi_{0k})_{k=1}^3$, refer to \textbf{Proof of Theorem~\ref{pf:NS_ONB}}. 

Letting $\theta$ vary from a reasonable range, we observe from Figure~\ref{fig:NS_ONB} that $\psi_{01}$ is always the level, $\psi_{02}$ the slope, and $\psi_{03}$ the curvature. Furthermore, with an increase in positive $\theta$ increases, the slope and the curvature change more drastically at short-maturity bonds; a decrease in negative $\theta$ pushes the changes of the slope and curvature towards long-term yields. Hence, just like the coefficients $(\beta_k)_{k=0}^2$ in Eq.~\eqref{eq:NSoriginal}, the coordinates in the new basis $(\psi_{0k})_{k=1}^3$ can still be interpreted as the loadings on the level/slope/curvature factors, if they truly are these factors. 

If we view $(\psi_{0k})_{k=1}^3$ as the factors with the most significant loadings among an extended orthonormal set $(\psi_{0k})_{k=1}^{\infty}$, we now proceed to test if $(\psi_{0k})_{k=1}^{\infty}$ is equivalent to the actual eigenfunctions $(\psi_k)_{k=1}^\infty$ up to permutations and changes of sign, i.e., whether the orthonormal basis underlying the Nelson-Siegel model is well-specified to match the actual principal component basis of the covariance kernel $G$.

\subsection{Covariance Estimation}
\begin{figure}[H]
    \centering
    \includegraphics[width=0.7\textwidth]{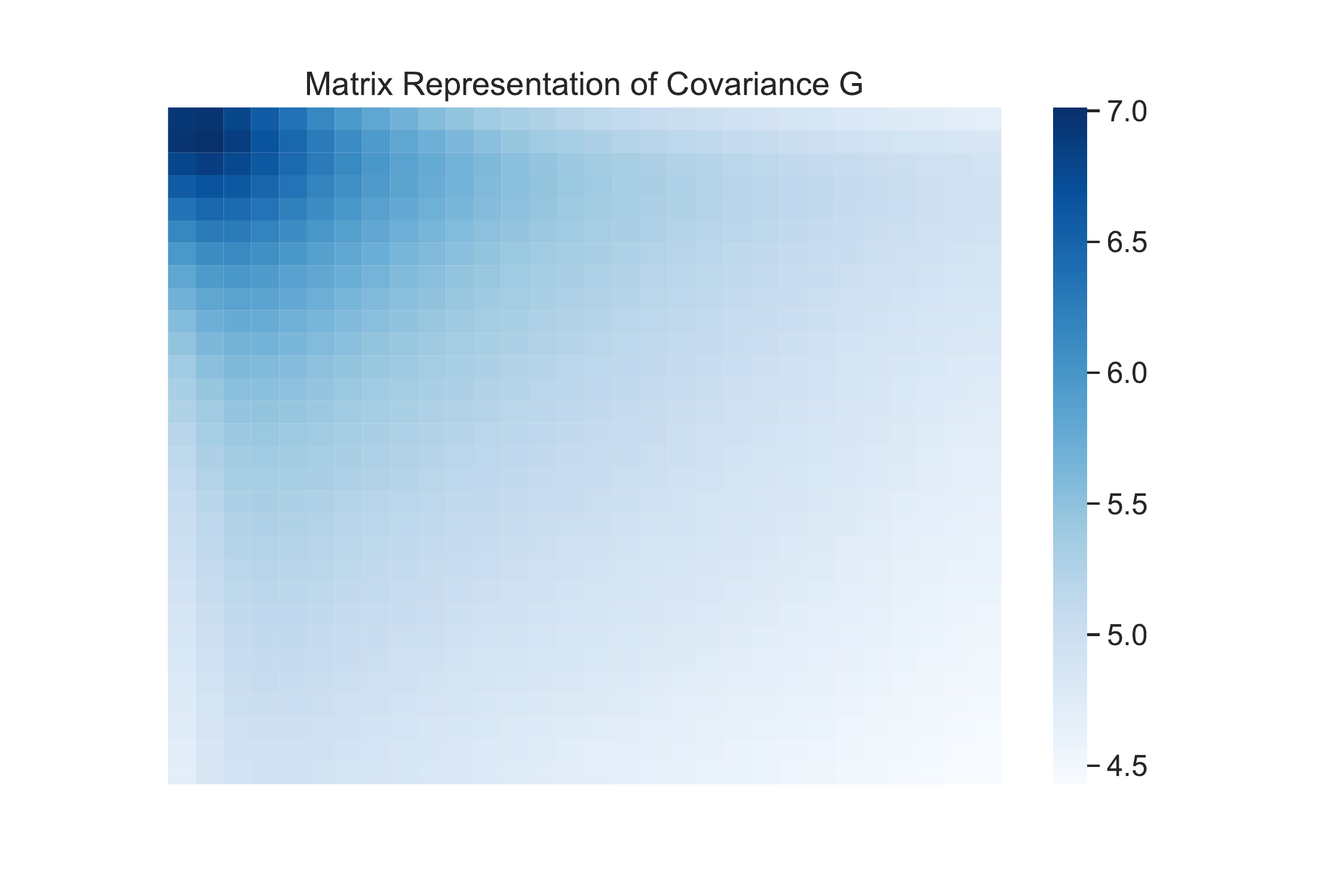}
    \caption{$\hat{G}$ as a low-rank sample covariance matrix on the entire data ($N=30$).}
    \label{fig:Ghat_all}
\end{figure}

We discretize $x$ and $x'$ on an $N\times N$ grid and estimate a rank-$N$ approximation of the bi-infinite covariance matrix in Eq.~\eqref{eq:covEstimate}. The entries of the low-rank sample covariance matrix are shown in Figure~\ref{fig:Ghat_all} for the entire data. The positive entries demonstrate the strong cross-sectional co-movements of yield curves over time. Such co-movements are much stronger on short-term bond yields as concentrated at the upper-left block of Figure~\ref{fig:Ghat_all}, which apparently have higher liquidity and respond to the market more effectively than long-term bond yields.

\begin{figure}[H]
    \centering
    \includegraphics[width=\textwidth]{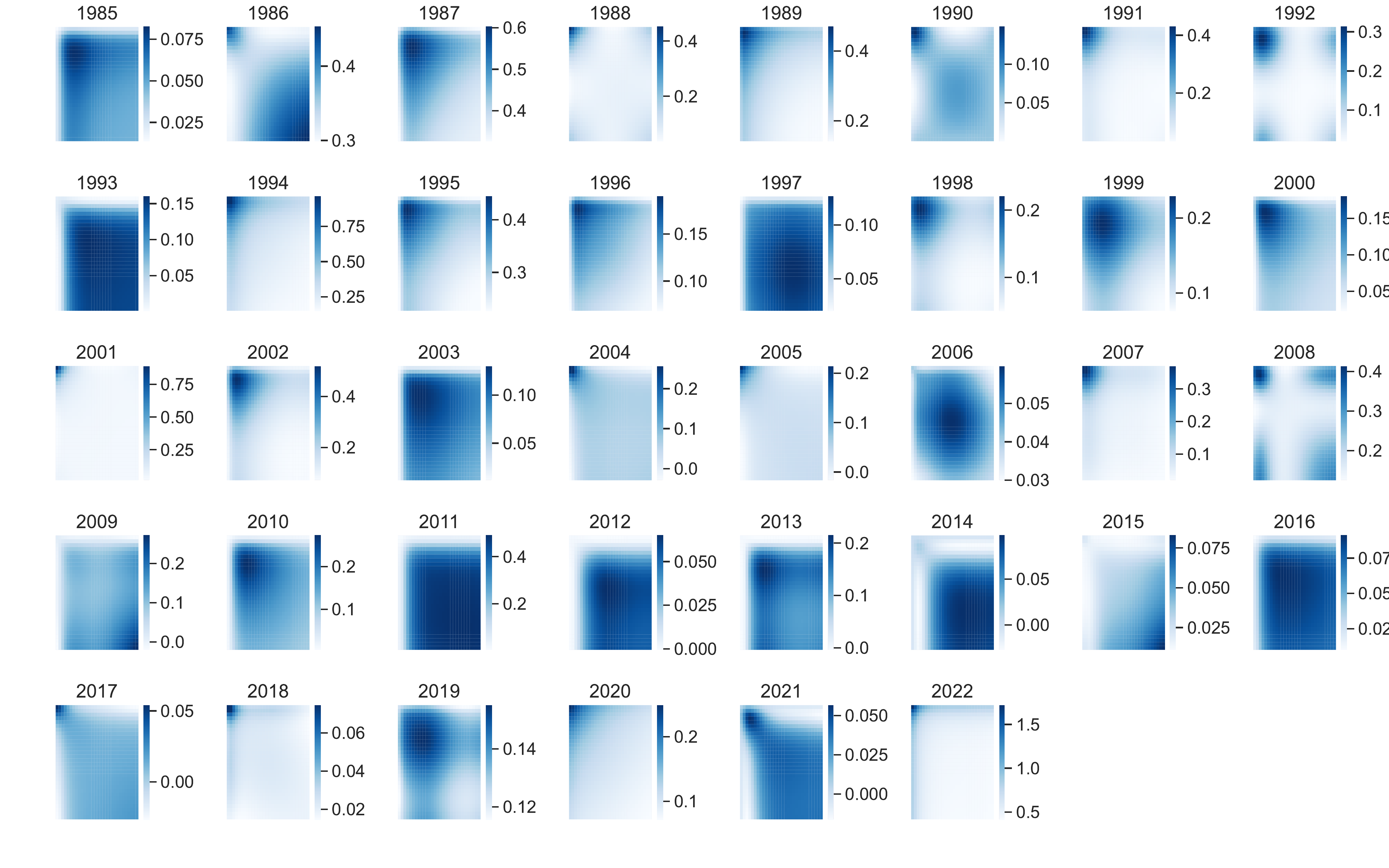}
    \caption{Implied $\hat{G}$ each year ($N=30$).}
    \label{fig:GhatYear}
\end{figure}

However, such a statement is only true when there are no significant frictions in the market. Notice the estimates $\hat{G}$ are heterogeneous in different years shown in Figure~\ref{fig:GhatYear}. Unlike \citet{hu2013noise}'s noise measure of liquidity and arbitrage capital that equally weighed the pricing errors of bond yields with different maturities, we argue that the cross-sectional difference in covariance is also indicative for different sources of liquidity. In 1986, for example, the short-term covariation and long-term covariation were both high, while the covariance between short-term yields and long-term ones stayed low, which indicates that the frictions that drove the illiquid yields might have some components uncorrelated with everyday frictions that mainly impacted the short-term yields, and further gave insights on the 1987 market crash. For another instance, a different but unique covariance structure was observed in both 1992 and 2008, representing the strong and correlated co-movements on both short-term and long-term yields, which could be impacted by factors from the external financial market which tilted the yield curves at their most extreme values, such as the exchange rate risk exposure in the 1992 UK currency crisis and the overall exposure to the global financial crisis in 2008.

\begin{figure}[H]
    \centering
    \includegraphics[width=\textwidth]{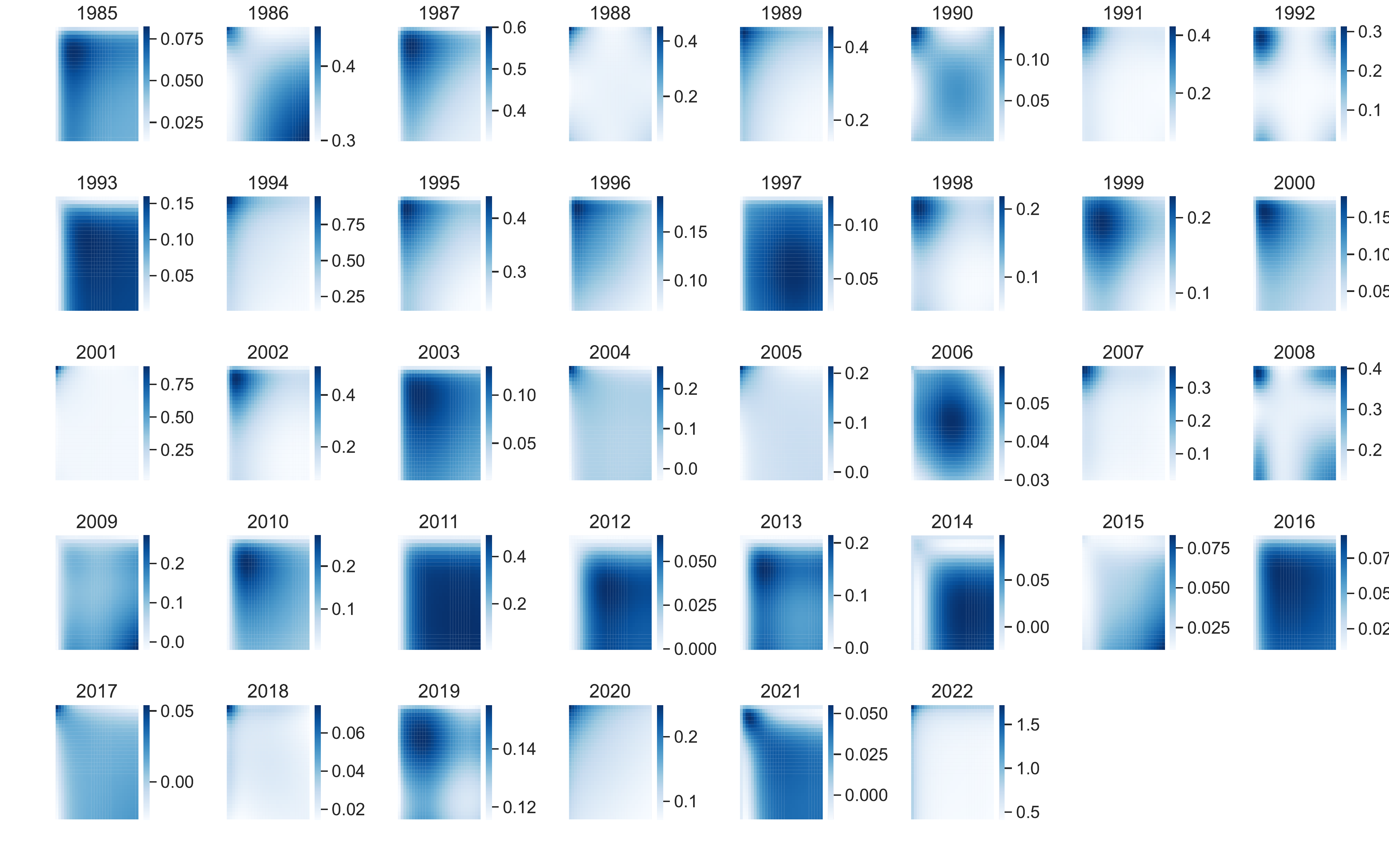}
    \caption{Yearly implied $\hat{G}$ fitted with linear B-splines and 30 uniform knots.}
    \label{fig:GhatYear_order1}
\end{figure}

\begin{figure}[H]
    \centering
    \includegraphics[width=\textwidth]{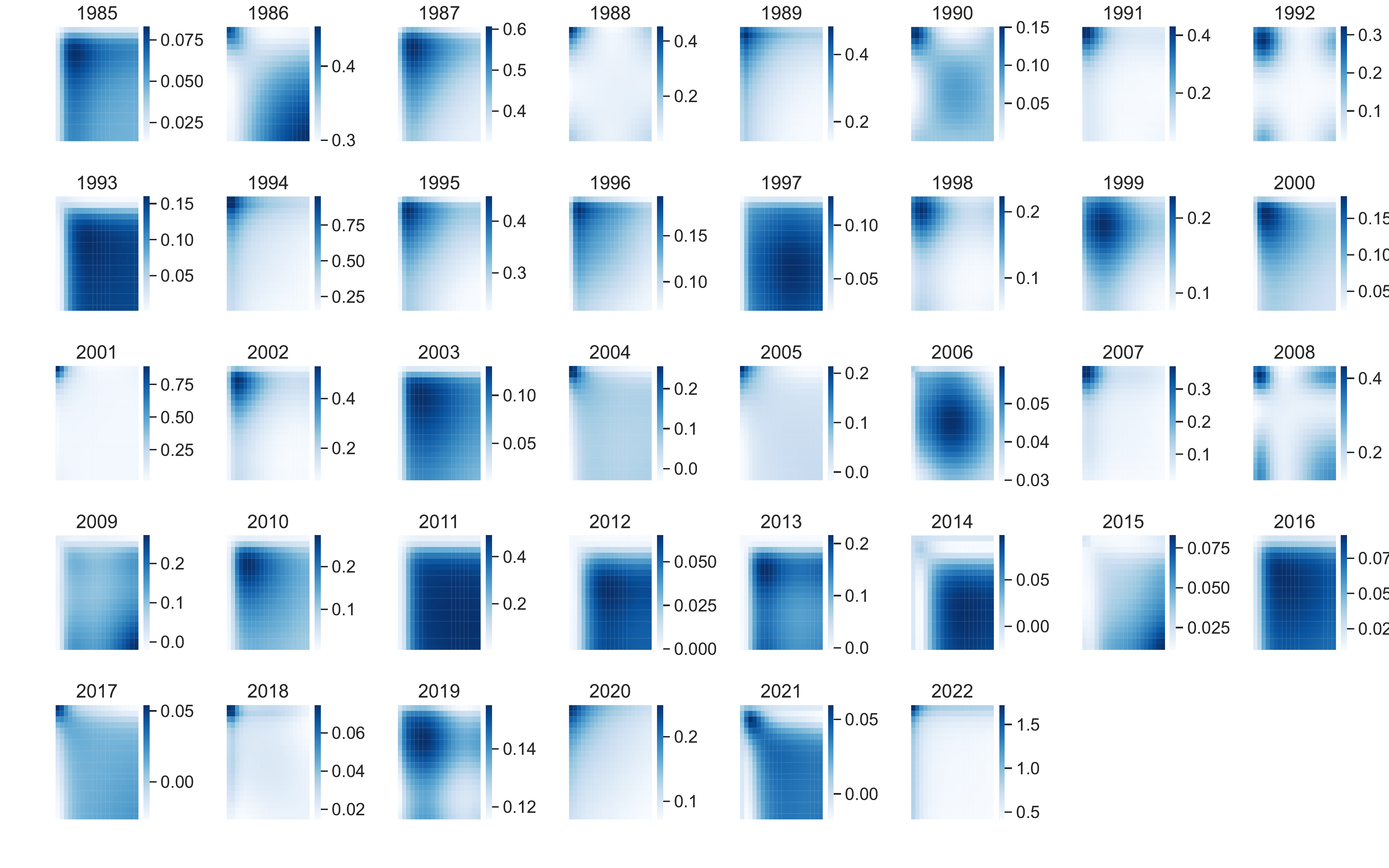}
    \caption{Yearly implied $\hat{G}$ fitted with cubic B-splines and 20 uniform knots.}
    \label{fig:GhatYear_knots20}
\end{figure}

To show the interpretations above are robust with respect to the choice of B-spline subspace and the number of uniform knots $N$, so that the projection in Eq.~\eqref{eq:Bspline} is not overfitting the yield curves, we demonstrate highly similar patterns revealed by Figure~\ref{fig:GhatYear_order1} using linear B-spline basis and by Figure~\ref{fig:GhatYear_knots20} where $N=20$. This also justifies the robustness of the estimation of covariance function $G$.

\subsection{Interpreting Functional Principal Components}\label{sec:interpretFPCA}
\begin{figure}[H]
    \centering
    \includegraphics[width=\textwidth]{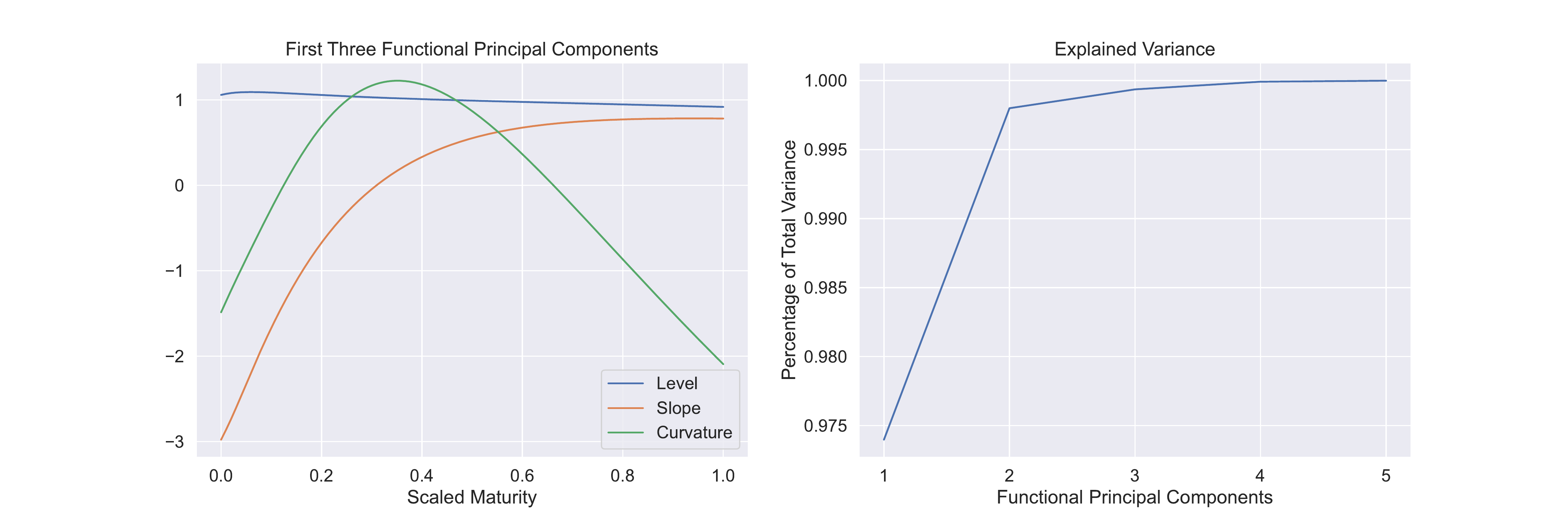}
    \caption{First three functional principal components of the entire dataset and explained variance ratios.}
    \label{fig:first3FPC_allData}
\end{figure}

Based on the covariance estimate $\hat{G}$, we first straightforwardly implement FPCA, the results given in Figure~\ref{fig:first3FPC_allData}. The shapes of the first three functional principal components and their explained variance are almost identical to the PCA results in Figure~\ref{fig:first3PC_allData}.

\begin{figure}[H]
    \centering
    \includegraphics[width=\textwidth]{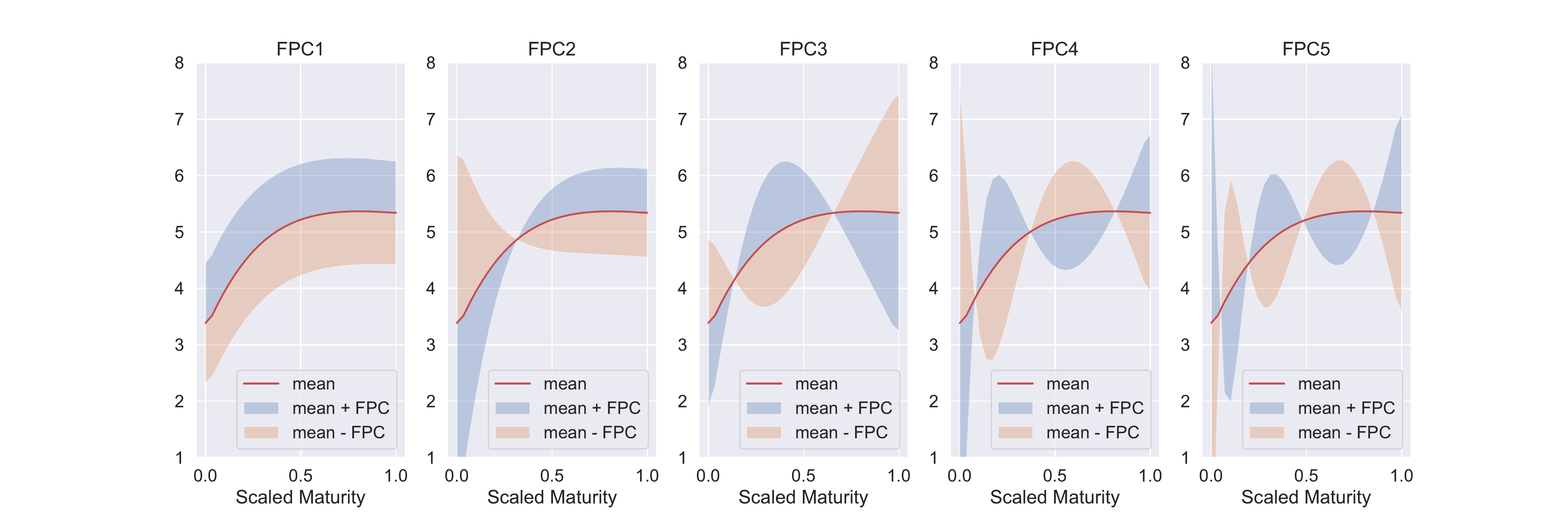}
    \caption{Effects of the functional principal components: blue shades represent the effects of positive shocks in the principal component, and orange shades represent those of negative shocks.}
    \label{fig:FPC_effects}
\end{figure}

To better illustrate the effects of the obtained functional principal components, we add and subtract the components to the mean function $\hat{m}$. We can then observe that the first component (FPC1) shows the parallel shift in the mean $\hat{m}$, the second component (FPC2) changes the slope of yield curves by tilting around one knot, the third (FPC3) embeds rich curvature by tilting the yield curve around two fixed knots, FPC4 tilts around 3 points, and so on so forth. Thus, similar interpretations on the level, slope, and curvature to PCA could be given to the functional principal components. Figure~\ref{fig:FPC_effects} offers an intuition of how the flexibility of yield curves can be traced back by their functional principal components. Formally, such intuition is a natural result of the following theorem.

\begin{theorem}\label{thm:DGP}
    If the data-generating process with identically distributed (can be dependent) zero-mean noise $\epsilon$ is
\be
y(\tau) = \beta_0 + \beta_1 \frac{1-e^{-\theta_1\tau}}{\theta_1\tau}
+ \beta_2 \left(
\frac{1-e^{-\theta_1\tau}}{\theta_1\tau} - e^{-\theta_1\tau}
\right)
+ \beta_3 \left(
\frac{1-e^{-\theta_2\tau}}{\theta_2\tau} - e^{-\theta_2\tau}
\right) + \epsilon,
\label{eq:DGP}\ee
then the number of zeros of the $k$-th functional principal component $|Ker(\psi_k)|\overset{p}{\to} k-1$.\footnote{Here we loosely use $Ker(f)$ to represent the set of zeros of a function $f$. It is not a subspace.}
\end{theorem}

Eq.~\eqref{eq:DGP} is also the \citet{svensson1994estimating} specification of yield curves on all maturities, also used in our data after 1980\footnote{Prior to 1980, the Nelson-Siegel model with fewer parameters was used to fit the yield curve, as there were not enough Treasury securities to fit the Svensson model, which has six parameters.}~\citep{gurkaynak2007us} and widely accepted as smoothed yields on hypothetical Treasury securities\footnote{Refer to the nominal yield curve (\url{https://www.federalreserve.gov/data/nominal-yield-curve.htm}).}. Regarding the parameters, $\theta_i=\frac{1}{\tau_i}$ for~$i=1,2$, and $(\beta_k)_{k=0}^3$ are the same as the Svensson model.

\begin{figure}[H]
    \centering
    \includegraphics[width=\textwidth]{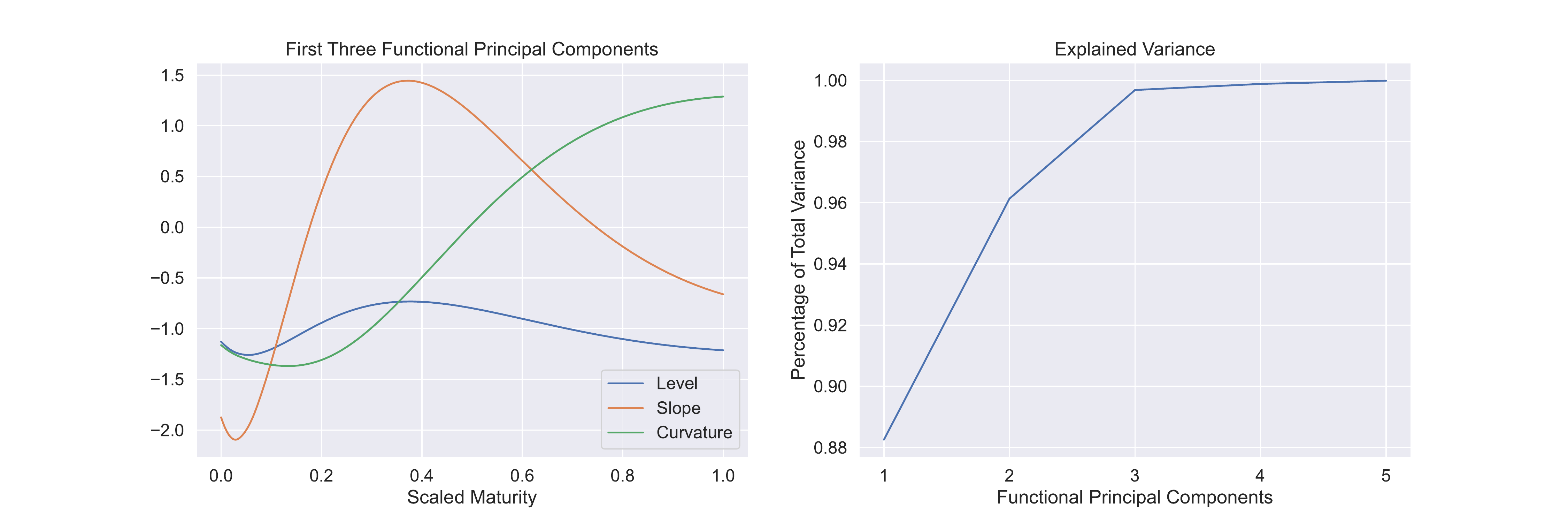}
    \caption{First three functional principal components in 2008 and explained variance ratios.}
    \label{fig:first3FPC_allData_2008}
\end{figure}

\textbf{Theorem~\ref{thm:DGP}} also implies that if $|Ker(\psi_k)|$ deviates a lot from $k-1$, either the regularity conditions imposed on $\epsilon$ fail to be satisfied, or the sample size is not large enough, or the data-generating process is misspecified. We argue that the former two scenarios are explicitly testable, both involving tests (whether asymptotic or not) on the distribution of $\epsilon$. So it would be possible that the model is misspecified after eliminating the former two scenarios. For instance, it might not be appropriate to name the corresponding functional principal components in Figure~\ref{fig:first3FPC_allData_2008} according to level, slope, and curvature anymore, and the fact that $|Ker(\psi_k)|\neq k-1$ for $k=2,3$ is evidence in favor of model misspecification.

\begin{figure}[H]
    \centering
    \includegraphics[width=\textwidth]{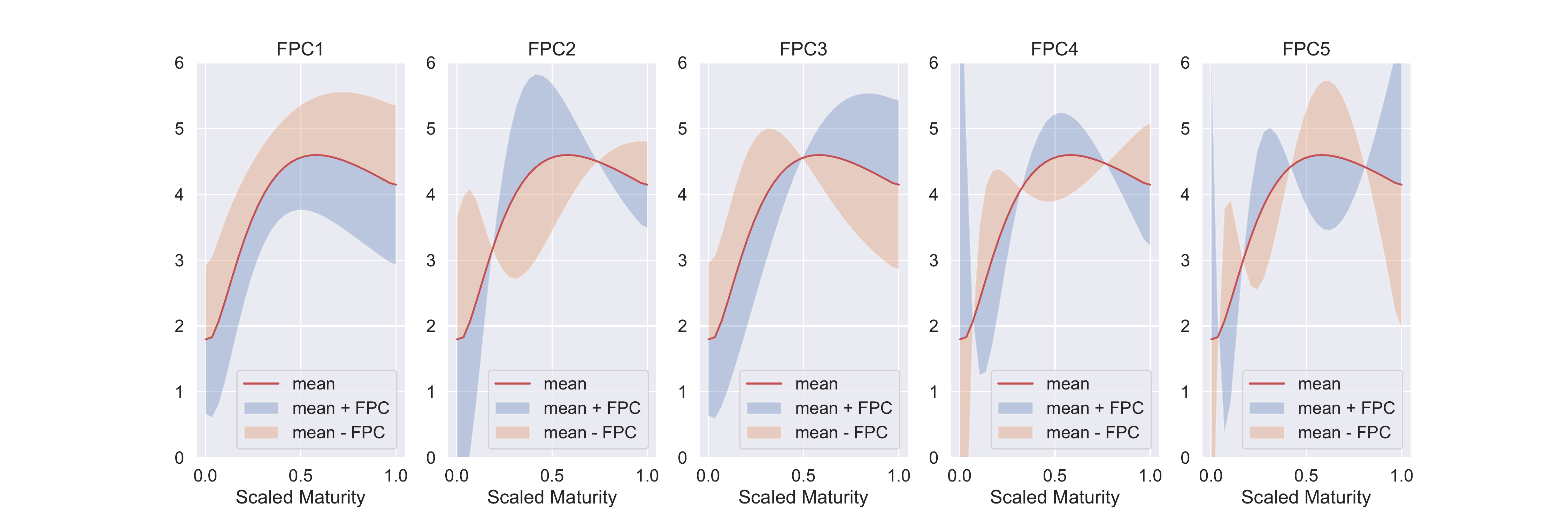}
    \caption{Effects of the functional principal components in 2008: $|Ker(\psi_2)|=2$ and $|Ker(\psi_3)|=1$.}
    \label{fig:FPC_effects_2008}
\end{figure}

\subsection{Test Statistic and $p$-value}\label{sec:testStat}
We now formally compute the test statistic $\hat{S}_n$ and simulate $\bar{S}_n$, the estimated asymptotic distribution of the test statistic under $H_0$, in the full sample. The hyper-parameter $\kappa_n=3$, since the eigenvalues $\lambda_k \ (k>3)$ are negligible under $H_0$, i.e., $(\psi_{0k})_{k=1}^3$ given in \textbf{Theorem}~\ref{thm:NS_ONB} already explains most variation in the yield curves.

\begin{figure}[H]
    \centering
    \includegraphics[width=\textwidth]{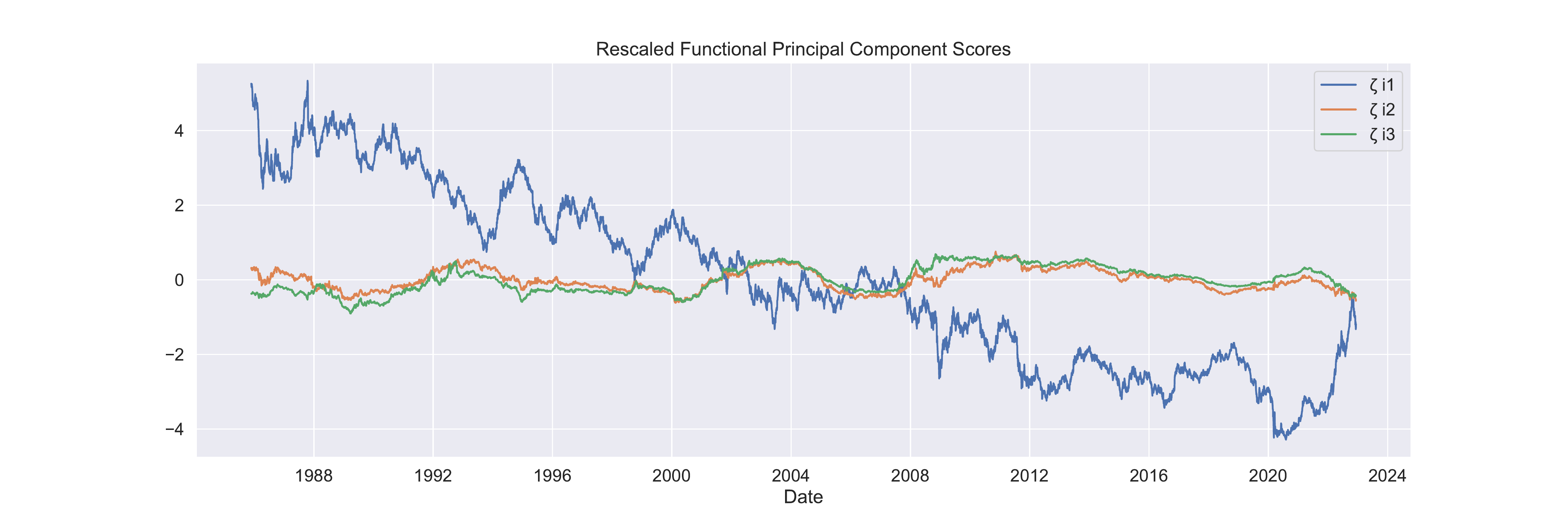}
    \caption{Rescaled functional principal component scores for $(\psi_{0k})_{k=1}^3$.}
    \label{fig:zeta}
\end{figure}

Comparing Figure~\ref{fig:zeta} and Figure~\ref{fig:timeSeriesYields}, we see the score $\zeta_{i1}$ on the first proposed eigenfunction $\psi_{01}$ captures the macroscopic variation in yield curves of all maturities, as a natural result of the definition of functional principal component analysis.

\begin{figure}[H]
    \centering
    \includegraphics[width=\textwidth]{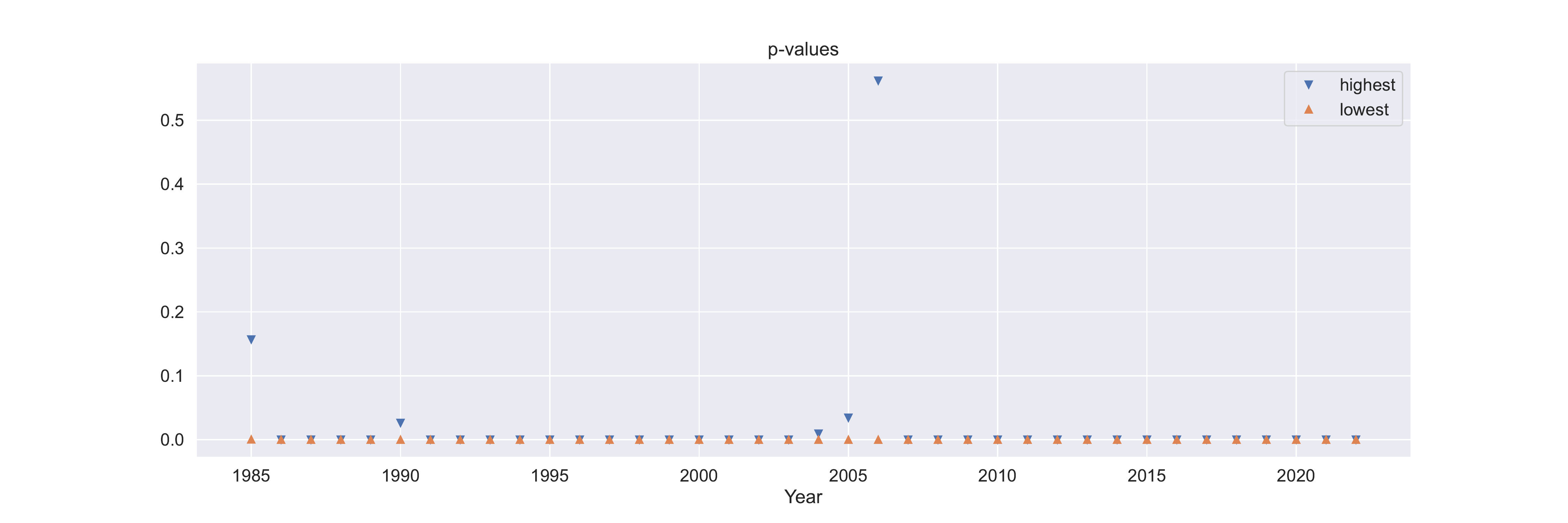}
    \caption{In the Song-Yang-Zhang test, the highest and lowest $p$-values attained for different parameter $\theta$ in each year.}
    \label{fig:pvalues_UT}
\end{figure}

Directly applying the test in \citet{song2022hypotheses} for a set of possible parameters $\theta\in\{\pm 0.001, \pm 0.01, \pm 0.1, 0.2, 0.5, 1\}$ for the Nelson-Siegel orthonormal basis, $H_0$ is rejected both on the entire sample and on almost each yearly data with $p << 0.05$ shown in Figure~\ref{fig:pvalues_UT}. This means that the Nelson-Siegel basis in general cannot be interpreted as the principal components, and there is strong evidence in the model being misspecified.

To further gain some insights on how confident we are when regarding the Nelson-Siegel orthonormal basis as the principal components, we loosen the above test statistic in \textbf{Theorem~\ref{thm:testStat}}.
\begin{theorem}\label{thm:testStat}
Modify the Song-Yang-Zhang test statistic in Eq.~\eqref{eq:testStat} to
$$
\hat{\tilde{S}}_n = n\sum\limits_{1\leq k\neq k'\leq\kappa_n} \hat{Z}_{kk'}^2,
$$
and the asymptotic distribution in Eq.~\eqref{eq:quadraticForm} to
$$
    \bar{\tilde{S}}_n = \sum\limits_{1\leq k\neq k'\leq\kappa_n} \hat{\lambda}_k\hat{\lambda}_{k'}\chi_{kk'}^2(1) +
    \sum\limits_{k=1}^{\kappa_n} \hat{\lambda}_k^2\left(
    \frac{1}{n}\sum\limits_{i=1}^n || \hat{\xi}_i^2 ||^2 -1
    \right)\chi_{kk}^2(1).
$$
Then the test is still consistent at asymptotic level of significance $\alpha$, i.e., under $H_0$,
$$
P(\hat{\tilde{S}}_n > \hat{\tilde{Q}}_{1-\alpha}) \to \alpha;
$$
and under $H_1$,
$$
P(\hat{\tilde{S}}_n > \hat{\tilde{Q}}_{1-\alpha}) \to 1,
$$
where $\hat{\tilde{Q}}_{1-\alpha}$ is the $100(1-\alpha)$-th percentile of $\bar{\tilde{S}}_n$.
\end{theorem}

Now we proceed to conduct the hypothesis test in \textbf{Theorem~\ref{thm:testStat}} on the full sample. The $p$-values reported in Table~\ref{tab:pvalues} suggest that the Nelson-Siegel basis $(\psi_{0k})_{k=1}^3$ is not highly sensitive enough to the parameter $\theta$ to influence the test conclusion within a reasonable range (e.g., $|\theta|\in[0.01, 0.2]$). Hence there is probably rich flexibility embedded by $\theta$ in the Nelson-Siegel model.
\begin{table}[H]
    \centering
    \begin{tabular}{c|c}
    \hline\hline
     $\theta$ & $p$-values \\
     \hline\hline
        $-0.1$ &  0.1944 \\
        $-0.01$ & 0.1946\\
        $-0.001$ & 0.0932\\
        0.001 & 0.0515\\
        0.01 & 0.1925\\
        0.1 & 0.1724\\
        0.2 & 0.1506\\
        0.5 & 0.0744\\
        1 & $0.0169 ^*$\\
         \hline\hline
    \end{tabular}
    \caption{$p$-values on full sample for different values of $\theta$.}
    \label{tab:pvalues}
\end{table}

\begin{figure}[H]
    \centering
    \includegraphics[width=\textwidth]{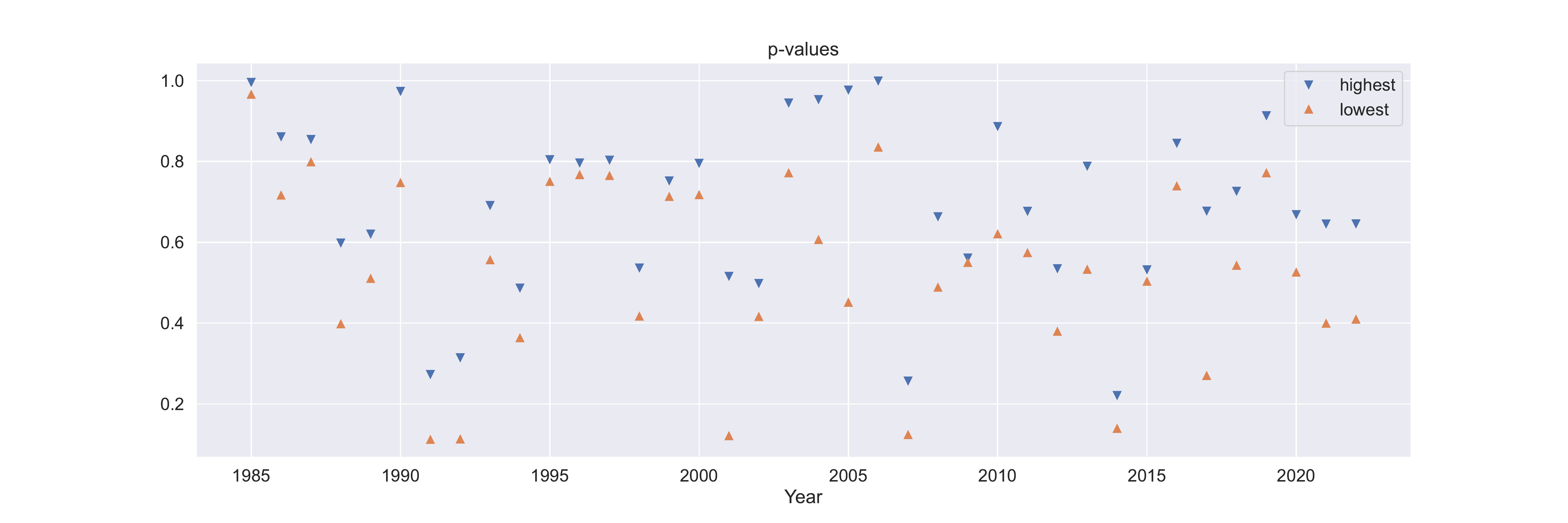}
    \caption{The highest and lowest $p$-values attained for different parameter $\theta$ in each year based on the test in \textbf{Theorem~\ref{thm:testStat}}.}
    \label{fig:pvalues}
\end{figure}

However, the heterogeneity of the test results each year in demonstrated by Figure~\ref{fig:pvalues}. We notice that when the highest $p$-values attained are low (so that there is evidence in favor of $H_1$, and the level-slope-curvature argument is suspicious),  there was a recession or some influential adjustments on the interest rate policies. In 2014, for instance, the Federal Reserve ceased its expansion in holding  longer-term securities through open market purchases, originally aiming at putting downward pressure on longer-term interest rates and supporting economic activities through making more accommodating financial conditions. 2007 marked that the Federal Reserve cut the target on short-term interest rate by half of a percentage point. The recession and RTC in 1991, and the UK currency crisis in 1992 causes explainable deviations to the reference model as well. These are all possible shocks that lead to interest rate model deviating from the level-slope-curvature component structure.

\section{Conclusions}
This paper investigates the model misspecification of interest rate term structure using functional principal component analysis. 
We first show by orthonormalization in the function space that the Nelson-Siegel model, as a standard reference model, can be interpreted as a three-factor interest rate term structure involving the level, the slope, and the curvature. This justifies the presence of the Nelson-Siegel orthonormal basis in the null hypothesis to test whether the actual data-driven principal components coincide with the level/slope/curvature factors, i.e., whether (and when) the three-factor model is misspecified.

Two notions have been used to determine whether two bases are the same. The stricter one in Section~\ref{sec:interpretFPCA} offers a non-rigorous way to test whether the functional principal components match the level, the slope, and the curvature in the exact order. We conclude from the entire sample that the components appear in the desired order, which is not the case for specific individual years. For instance, in 2008, a curvature factor potentially explained more variance than a slope factor. 
The same conclusion is drawn based on the other less strict notion in Section~\ref{sec:testStat}, where permutations and sign changes in the basis are accepted. To sum up, this paper characterizes the heterogeneity of misspecification in the interest rate term structure when rare events or relevant shocks in policies occur, even if the model is accepted overall.

Future research could generalize the proposed methodology to test the reference models other than the Nelson-Siegel model, which lacks financial intuition compared to other affine models such as \citet{vasicek1977equilibrium}, and to include calibration from the actual market data in bond yields rather than merely hypothetical securities.

\newpage
\onehalfspacing
\setlength\bibsep{0pt}
\bibliographystyle{aea}
\bibliography{literature}

\clearpage

\doublespacing
\clearpage

\section*{Appendix A. Proofs} \label{app:proofs}
\addcontentsline{toc}{section}{Appendix A}

\begin{proofTheorem}\label{pf:NS_ONB}
\begin{proof}
Consider the following rescaling of $\tau$
$$
\tilde{\tau} = \frac{\tau - \ubar{\tau}}{\bar{\tau} - \ubar{\tau}} \in [0,1].
$$
Then Eq.~\eqref{eq:NSoriginal} becomes
    $$
\hat{y}(\tilde{\tau}\mid \Theta) = \sum\limits_{k=1}^3 \alpha_k \Psi_{0k}(\tilde{\tau}) \in L^2([0,1],d\xi),
$$
where the basis functions
\bea 
\Psi_{01}(\tilde{\tau}) & = & 1, \\
\Psi_{02}(\tilde{\tau}) & = & e^{-\theta\tau} = \exp \left\{ -\theta[(\bar{\tau} - \ubar{\tau})\tilde{\tau}+\ubar{\tau}] \right\}
=\exp \left\{ -(A\tilde{\tau}+B) \right\}, \\
\Psi_{03}(\tilde{\tau}) & = & \frac{1-e^{-\theta\tau}}{\theta\tau}
=\frac{1-\exp \left\{ -\theta[(\bar{\tau} - \ubar{\tau})\tilde{\tau}+\ubar{\tau}] \right\}}{\theta[(\bar{\tau} - \ubar{\tau})\tilde{\tau}+\ubar{\tau}] }
=\frac{1-\exp \left\{ -(A\tilde{\tau}+B) \right\}}{A\tilde{\tau}+B}, 
\eea
with
\bea
A & = & \theta(\bar{\tau} - \ubar{\tau}), \\
B & = & \theta \ubar{\tau},
\eea
and the coordinates 
$$
(\alpha_{1},\alpha_{2},\alpha_{3})' = (\beta_0,-\beta_2,\beta_1+\beta_2)'.
$$

Now we follow the Gram–Schmidt process to obtain an orthonormal basis $(\psi_{0k})_{k=1}^3$ for the subspace $Span(\Psi_{0k})_{k=1}^3$.
Start with $\psi_{01}=\Psi_{01}=1$, which satisfies $||\psi_{01}||^2 = \int\limits_0^1 d\xi = 1.$
Consider the projection of $\Psi_{02}$ onto $\psi_{01}$
$$
\gP_{\psi_{01}}\Psi_{02} = \frac{\langle \psi_{01}, \Psi_{02} \rangle}{\langle \psi_{01}, \psi_{01} \rangle}\psi_{01} = \int\limits_0^1 
\exp \left\{ -(A\xi+B) \right\} d\xi = \frac{e^{-B}-e^{-A-B}}{A} =: D.
$$
Then
$$
\tilde{\psi}_{02} = \Psi_{02} - \gP_{\psi_{01}}\Psi_{02}
= \exp \left\{ -(A\tilde{\tau}+B) \right\} - D,
$$
with
\bea
||\tilde{\psi}_{02}||^2 &= & \left\langle \tilde{\psi}_{02}, \tilde{\psi}_{02} \right\rangle \\
&= &  \frac{1}{2A}(e^{-2B}-e^{-2(A+B)}) - \frac{2D}{A}(e^{-B}-e^{-A-B}) + D^2 \\
&= & \frac{1}{2A}(e^{-2B}-e^{-2(A+B)}) - D^2.
\eea
Thus, 
$$
{\psi}_{02} = \frac{\tilde{\psi}_{02}}{||\tilde{\psi}_{02}||}.
$$
The projection of $\Psi_{03}$ onto $\psi_{01}$ is
\bea
\gP_{\psi_{01}}\Psi_{03} &=& \frac{\langle \psi_{01}, \Psi_{03} \rangle}{\langle \psi_{01}, \psi_{01} \rangle}\psi_{01}\\
&=& \int\limits_0^1 
\frac{1-\exp \left\{ -(A\xi+B) \right\}}{A\xi+B} d\xi \\
&=& \frac{\log|A+B|-\log|B|-Ei(-A-B)+Ei(-B)}{A} \\
&=&\frac{\log|A+B|-\log|B|}{A} - \frac{Ei(-A-B)-Ei(-B)}{A} := F-H,
\eea
where $Ei$ is the exponential integral.
The projection of $\Psi_{03}$ onto $\tilde{\psi}_{02}$ is
$$
\gP_{\tilde{\psi}_{02}}\Psi_{03} = \frac{\langle \tilde{\psi}_{02}, \Psi_{03} \rangle}{\langle \tilde{\psi}_{02}, \tilde{\psi}_{02} \rangle}\tilde{\psi}_{02}
= \frac{\tilde{\psi}_{02}}{||\tilde{\psi}_{02}||^2} 
\left\langle \tilde{\psi}_{02}, \Psi_{03} \right\rangle,
$$
where
\bea
\left\langle \tilde{\psi}_{02}, \Psi_{03} \right\rangle &=&
\int\limits_0^1 \left(\exp \left\{ -(A\xi+B) \right\} - D \right)
\frac{1-\exp \left\{ -(A\xi+B) \right\}}{A\xi+B} d\xi \\
&=& \int\limits_0^1 
\frac{\exp \left\{ -(A\xi+B) \right\} -\exp \left\{ -2(A\xi+B) \right\}}{A\xi+B} d\xi
- D\int\limits_0^1 
\frac{1-\exp \left\{ -(A\xi+B) \right\}}{A\xi+B} d\xi\\
&=& \frac{Ei(-A-B)-Ei(-B)-Ei(-2A-2B)+Ei(-2B)}{A} -D(F-H) \\
&=& H - \frac{Ei(-2A-2B)-Ei(-2B)}{A} -D(F-H) \\
&=& (D+1)H - DF - \frac{Ei(-2A-2B)-Ei(-2B)}{A} =: J.
\eea
Then
\bea
\tilde{\psi}_{03} &=& \Psi_{03} - \gP_{\psi_{01}}\Psi_{03} - \gP_{\tilde{\psi}_{02}}\Psi_{03} \\
&=& \frac{1-\exp \left\{ -(A\tilde{\tau}+B) \right\}}{A\tilde{\tau}+B} - F + H  - \frac{J\tilde{\psi}_{02}}{||\tilde{\psi}_{02}||^2} \\
&=& \frac{1-\exp \left\{ -(A\tilde{\tau}+B) \right\}} {A\tilde{\tau}+B} + K\exp \left\{ -(A\tilde{\tau}+B) \right\} +L, 
\eea
where
\bea
K &:=& -\frac{J}{||\tilde{\psi}_{02}||^2}, \\
L &:=& -F + H +\frac{JD}{||\tilde{\psi}_{02}||^2}.
\eea
Finally,
\bea
||\tilde{\psi}_{03}||^2 &= & \left\langle \tilde{\psi}_{03}, \tilde{\psi}_{03} \right\rangle \\
&=& \int\limits_0^1 \left(
\frac{1-\exp \left\{ -(A\xi+B) \right\}} {A\xi+B}
\right)^2 d\xi
+ K^2 \int\limits_0^1 \exp \left\{ -2(A\xi+B) \right\} d\xi
+ L^2 \\
&& + 2K\int\limits_0^1 
\frac{\exp \left\{ -(A\xi+B) \right\} -\exp \left\{ -2(A\xi+B) \right\}}{A\xi+B} d\xi \\
&& + 2L \int\limits_0^1 
\frac{1-\exp \left\{ -(A\xi+B) \right\}} {A\xi+B}
 d\xi \\
 && + 2KL \int\limits_0^1 \exp \left\{ -(A\xi+B) \right\} d\xi \\
 &=& \frac{1}{B(A+B)} + \frac{2e^{-A-B}}{A(A+B)} + \frac{2Ei(-A-B)}{A} - \frac{2e^{-B}}{AB} -\frac{2Ei(-B)}{A}
- \frac{e^{-2A-2B}}{A(A+B)} \\
&& - \frac{2Ei(-2A-2B)}{A} + \frac{e^{-2B}}{AB} + \frac{2Ei(-2B)}{A}  + K^2 \frac{e^{-2B} - e^{-2(A+B)}}{2A} + L^2 \\
&& + 2K \left(H - \frac{Ei(-2A-2B)-Ei(-2B)}{A}\right) +2L(F-H) + 2KLD,
\eea
and
$$
{\psi}_{03} = \frac{\tilde{\psi}_{03}}{||\tilde{\psi}_{03}||}.
$$
So far, we have recovered an orthonormal basis $(\psi_{0k})_{k=1}^3$.

\end{proof}
\end{proofTheorem}

\begin{proofTheorem}\label{pf:k-1zeros}
\begin{proof}
For notational simplicity, write
\bea
\Psi_{02}^\theta(\tau) &=& e^{-\theta\tau}, \\
\Psi_{03}^\theta(\tau) & = & \frac{1-e^{-\theta\tau}}{\theta\tau},
\eea
for $\theta\in\{\theta_1, \theta_2\}$.
Then Eq.~\eqref{eq:DGP} becomes
\bean
y &=& \beta_0 + \beta_1 \Psi_{03}^{\theta_1}
+ \beta_2 (\Psi_{03}^{\theta_1} - \Psi_{02}^{\theta_1})
+ \beta_3 (\Psi_{03}^{\theta_2} - \Psi_{02}^{\theta_2}) + \epsilon \nonumber\\
&=& \beta_0 + (\beta_1+\beta_2)\Psi_{03}^{\theta_1} - \beta_2\Psi_{02}^{\theta_1} + \beta_3\Psi_{03}^{\theta_2} - \beta_3\Psi_{02}^{\theta_2} + \epsilon.
\label{eq:DGPsimplified}\eean

Compute the first-order derivatives
\bea
\frac{\partial}{\partial \tau}\Psi_{03}^{\theta}
&=& -\theta \left(
\frac{1-e^{-\theta\tau}}{\theta^2\tau^2} -\frac{ e^{-\theta\tau}}{\theta\tau} \right) \\
&=& -\frac{1}{\tau}\left(
\frac{1-e^{-\theta\tau}}{\theta\tau} - e^{-\theta\tau}
\right) \\
&=& -\frac{1}{\tau}(\Psi_{03}^{\theta} - \Psi_{02}^{\theta})
\eea
and
\bea
\frac{\partial}{\partial \tau}\Psi_{02}^{\theta} &=& -\theta\Psi_{02}^{\theta}.
\eea
The second-order derivatives are hence
\bea
\frac{\partial^2}{\partial \tau^2}\Psi_{03}^{\theta}
&=& -\frac{1}{\tau^2} \left(
\tau\frac{\partial}{\partial \tau}\Psi_{03}^{\theta} - \Psi_{03}^{\theta} - \tau\frac{\partial}{\partial \tau}\Psi_{02}^{\theta} + \Psi_{02}^{\theta}
\right) \\
&=& -\frac{1}{\tau^2} \left(
 \Psi_{02}^{\theta} - 2 \Psi_{03}^{\theta} + (\theta\tau+1)  \Psi_{02}^{\theta}
\right) \\
&=& \frac{1}{\tau^2} \left(
2 \Psi_{03}^{\theta} - (\theta\tau+2)  \Psi_{02}^{\theta}
\right)
\eea
and
$$
\frac{\partial^2}{\partial \tau^2}\Psi_{02}^{\theta}
= -\theta \frac{\partial}{\partial \tau}\Psi_{02}^{\theta} = \theta^2\Psi_{02}^{\theta}.
$$
Under some regularity conditions on $\epsilon$, we can take the derivatives in Eq.~\eqref{eq:DGPsimplified}:
$$
\frac{\partial y}{\partial \tau} = -\frac{\beta_1+\beta_2}{\tau} (\Psi_{03}^{\theta_1} - \Psi_{02}^{\theta_1}) + \beta_2\theta_1\Psi_{02}^{\theta_1} 
 -\frac{\beta_3}{\tau} (\Psi_{03}^{\theta_2} -\Psi_{02}^{\theta_2}) + \beta_3\theta_2\Psi_{02}^{\theta_2}
$$
and
$$
\frac{\partial^2 y}{\partial \tau^2} = \frac{\beta_1+\beta_2}{\tau^2} \left[
2 \Psi_{03}^{\theta_1} - (\theta_1\tau+2)  \Psi_{02}^{\theta_1} \right] - \beta_2\theta_1^2\Psi_{02}^{\theta_1} 
 + \frac{\beta_3}{\tau^2} \left[
2 \Psi_{03}^{\theta_2} - (\theta_2\tau+2)  \Psi_{02}^{\theta_2} \right] - \beta_3\theta_2^2\Psi_{02}^{\theta_2}.
$$

Impose the ansatz that there exist 
\bi
\item $p\in C^1[\ubar{\tau},\bar{\tau}]$,
\item $q\in C^0[\ubar{\tau},\bar{\tau}]$, 
\item $r\in C^0(0,\infty)$ s.t. $r(\lambda_k)\uparrow\infty$,
\ei
satisfying
$$
p\frac{\partial^2 y}{\partial \tau^2} + p'\frac{\partial y}{\partial \tau} + qy = -r y,
$$
or
\be
\frac{\partial}{\partial \tau} \left(p\frac{\partial y}{\partial \tau}\right) + qy = -ry.
\label{eq:sturm}\ee
The ansatz could be verified by matching the coefficients.
Then Eq.~\eqref{eq:sturm} with well-defined boundary conditions constitutes a regular Sturm–Liouville problem. 

According to the Sturm–Liouville theory, the $k$-th fundamental solution corresponding to the eigenvalue $\lambda_k$ is the unique eigenfunction $\psi_k$ with exactly $k-1$ zeros in $(\ubar{\tau},\bar{\tau})$ (in the probability limit under some regularity conditions for $\epsilon$). Thus,
$$
|Ker(\psi_k)|\overset{p}{\to} k-1.
$$

A special case of Sturm–Liouville theory when $\psi_k$'s are real polynomials is presented in \citet{hou2021new}, which also applies for many proposed $\psi_{0k}$ (e.g., in B-spline basis as the case in this paper, in monomial basis, or even in the orthonormal basis in \textbf{Theorem~\ref{thm:NS_ONB}} up to the errors of Taylor expansion of the exponential).

\end{proof}
\end{proofTheorem}

\begin{proofTheorem}\label{pf:testConsistency}
\begin{proof}
Let
$$
\mY_i = \left(\zeta_{ik}\zeta_{ik'} - \lambda_k\delta_{kk'}
\right)_{k,k'=1}^\infty,
$$
where $\delta$ is the Kronecker delta. According to \citet{song2022hypotheses}, the Hilbert central limit theorem yields
$$
\frac{1}{\sqrt{n}}\sum\limits_{i=1}^n \mY_i \overset{d}{\to} \mN\sim \gN(\vzero, \mC_\mY)
$$
with the covariance function
$$
\mC_\mY(\ve_{kk}) = \lambda_k^2(\E\xi_i^4 - 1)\ve_{kk}
$$
and
$$
\mC_\mY(\ve_{kk'}) = \lambda_k\lambda_{k'}\ve_{kk'}.
$$
Hence
$$
\frac{\mN_{kk}^2}{\lambda_k^2(\E\xi_i^4 - 1)},
\frac{\mN_{kk'}^2}{\lambda_k\lambda_{k'}} \sim \chi^2(1).
$$
Then by continuous mapping theorem, the Frobenius norm
\bean
\left|\left| \frac{1}{\sqrt{n}}\sum\limits_{i=1}^n \mY_i \right|\right|_F^2 &\overset{d}{\to}& \left|\left| \mN \right|\right|_F^2 \nonumber\\
&=& \sum\limits_{1\leq k\neq k'<\infty} {\lambda}_k{\lambda}_{k'}\chi_{kk'}^2(1) +
    \sum\limits_{k=1}^{\infty} {\lambda}_k^2\left(
    \E\xi_i^4 - 1
    \right)\chi_{kk}^2(1). \label{eq:adjustedTest}
\eean
Clearly, under $H_0$,
\be
 n\sum\limits_{1\leq k, k' < \infty} {Z}_{kk'}^2 = \left|\left| \frac{1}{\sqrt{n}}\sum\limits_{i=1}^n \mY_i \right|\right|_F^2.
\label{eq:adjustedTestStat} \ee
In the infeasible Gaussian quadratic form in Eq.~\eqref{eq:adjustedTest}, replace $\lambda$ and $\E\xi_i^4$ by their sample estimates and restrict the summation by the finite number $\kappa_n$. Similarly, replace the left-hand side in Eq.~\eqref{eq:adjustedTestStat} by $\hat{\tilde{S}}_n$.
The rest of the proof follows in a straightforward manner as \citet{song2022hypotheses}.

\end{proof}
\end{proofTheorem}

\clearpage

\section*{Appendix B. Robustness} \label{app:robust}
\addcontentsline{toc}{section}{Appendix B}

We demonstrate the robustness of our interpretations of Figure~\ref{fig:GhatYear} by showing its independence on the choice of $N$ and $p$ in Eq.~\eqref{eq:Bspline} as long as the yield curves are not overfit.

The functional principal components still exhibit similar properties as in Figure~\ref{fig:first3FPC_allData}.
\begin{figure}[H]
    \centering
    \includegraphics[width=\textwidth]{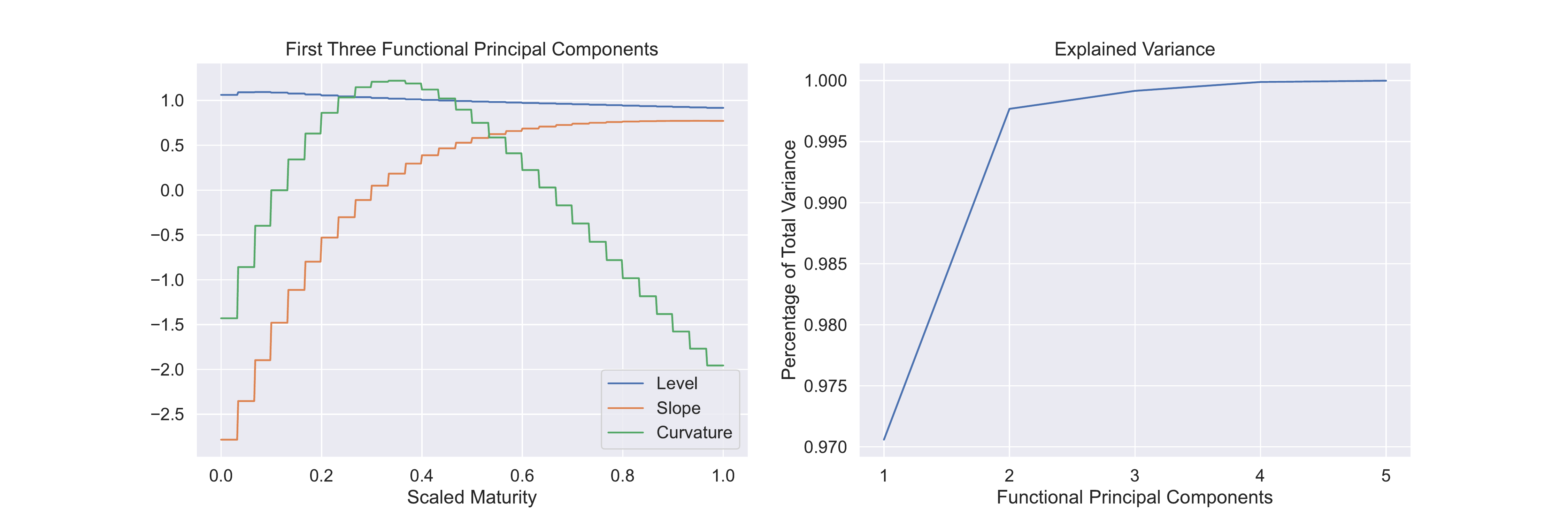}
    \caption{First three functional principal components with $\hat{G}$ fitted with linear B-splines and 30 uniform knots.}
    \label{fig:first3FPC_allData_linear}
\end{figure}

The number of knots is an even minor issue to our results. There is almost no discernible difference in the FPCA results using fewer uniform knots, so we omit the plot here.

\end{document}